\documentclass[english,prb,notitlepage,twocolumn]{revtex4-1}
\usepackage[T1]{fontenc}
\usepackage[latin9]{inputenc}
\usepackage{babel}
\usepackage{float}
\usepackage{amsmath}
\usepackage{graphicx}
\usepackage{xcolor}
\usepackage[unicode=true,pdfusetitle,
 bookmarks=true,bookmarksnumbered=false,bookmarksopen=false,
 breaklinks=false,pdfborder={0 0 1},backref=false,colorlinks=false]
 {hyperref}
 

\usepackage{babel}
\begin{document}
\title{Accuracy of ghost-rotationally-invariant slave-boson theory for multiorbital Hubbard models and realistic materials}
\author{Tsung-Han Lee$^{1,2}$, Corey Melnick$^{5}$, Ran Adler$^{1}$, Nicola Lanat\`a$^{3,4}$,
Gabriel Kotliar$^{1,5}$}
\affiliation{$^{1}$Physics and Astronomy Department, Rutgers University, Piscataway,
New Jersey 08854, USA}
\affiliation{$^{2}$Department of Physics, National Chung Cheng University, Chiayi 62102, Taiwan}
\affiliation{$^{3}$School of Physics and Astronomy, Rochester Institute of Technology,
84 Lomb Memorial Drive, Rochester, New York 14623, USA}
\affiliation{$^{4}$Center for Computational Quantum Physics, Flatiron Institute,
New York, New York 10010, USA}
\affiliation{$^{5}$Condensed Matter Physics and Materials Science Department,
Brookhaven National Laboratory, Upton, New York 11973, USA}
\begin{abstract}
We assess the accuracy of the ghost-rotationally-invariant slave-boson (g-RISB) theory in multiorbital systems by applying it to both the three-orbital degenerate Hubbard model and a realistic Sr$_2$RuO$_4$ model extracted from first principle simulations, and comparing the results to those obtained using the dynamical mean-field theory (DMFT). Our findings indicate that g-RISB's accuracy can be systematically improved toward the exact DMFT limit
%(solved with CTQMC)
in infinite dimensional multiorbital models by increasing the number of ghost orbitals. This allows for a more precise description of aspects of  Hund metal physics and Mott physics compared to the original RISB approach. We also demonstrate that g-RISB reliably captures the quasiparticle weights, Fermi surface, and {\color{black} low-energy} spectral function for the realistic Sr$_2$RuO$_4$ model compared to DMFT. %{\color{black}while the high energy Hubbard bands are approximated by coherent bands.}
{\color{black} Moreover, we showcase the potential of using the density matrix renormalization group method as an impurity solver within the g-RISB framework to study systems with a larger number of ghost orbitals.}
%Moreover, we explored a comparison between the two quantum embedding methods, solving the DMFT equations with exact-diagonalization (ED) as an impurity solver and the g-RISB equations with ED and density matrix renormalization group as an impurity solver.
These results show the potential of g-RISB as a reliable tool for simulating correlated materials. The connection between the g-RISB and DMFT self-energy is also discussed.

%\vspace{1.2cm}
\end{abstract}
\maketitle

\section{Introduction}

The rotationally-invariant slave-boson (RISB) mean-field theory~\citep{Lecherman_2007,Isidori_RISB_SC_2009}, Gutzwiller approximation~\citep{Gutzwiller1,Gutzwiller2,Metzner_Vollhardt_1989,Bunemann_mulorb_GA,Bunemann_2007,Fabrizio_SC,Lanata_2008}, and other slave-particle mean-field approaches are efficient methodologies for studying the strong correlation effects in multiorbital Hubbard models~\citep{Kotliar1986,Li1991,Medici2005,Matteo_SS_2023,Yu2012,Serge_rotor_2002,Serge_rotor_2004,Georgescu2015}.
These approaches effectively capture the fundamental orbital-selective Mott physics and Hund's metallicity in multiorbital systems~\citep{Koga_OSMT,Medici2005,Georges_Hunds_review,Medici_Janus_PRL,
HauleHunds,Fanfarillo_2015,Isidori_2019_PRL}, exhibiting qualitative agreement with the more computationally demanding dynamical mean-field theory (DMFT)~\citep{DMFT_RMP_1996},
and have been widely applied to investigate realistic materials, such as iron-based superconductors and heavy-fermion systems~\citep{Deng_2009,Lanata_2012,Lanata_2013_Ce,Lanata_2015_PRX,Lanata_2017_PRL,Lanata2019_NPJ,Christoph_2011,Lechermann_2018,Medici_2014}. 
However, despite their success, the accuracy of slave-particle mean-field approaches and the Gutzwiller approximation is not always sufficient. For example,  these approaches tend to overestimate the critical Coulomb interaction for Mott transitions due to the lack of descriptions of the charge fluctuation in the Mott insulating phase~\citep{gRISB_2021}. Hence, they require the use of larger Coulomb interactions to achieve experimentally observed Mott insulating behavior in transition metal compounds~\citep{Lanata2019_NPJ}.
%it is known that these approaches overestimate the Mott transition in Hubbard models \citep{Huang_Dai_2012}.
%Hence, larger Coulomb interactions have to be utilized to obtain experimental observed Mott insulating behavior in a series of transition metal compounds \citep{Lanata2019_NPJ}. 
Additionally, recent studies on Sr$_2$RuO$_4$ indicate that RISB significantly underestimates the effective mass of the electrons~\citep{Facio_2018_PRB,Lechermann_2019_SRO}. Therefore, it is desirable to have a systematic route to improve the accuracy of RISB. % with some sacrifice of computational efficiency.

Recently, the ghost-rotationally-invariant slave-boson (g-RISB) approach has been developed to improve the accuracy of the original RISB method~\citep{gRISB_2017,gRISB_2021,gRISB_2022}.
%Similar to the original RISB, the g-RISB theory maps the Hubbard model to an embedding impurity model and an auxiliary single-particle Hamiltonian. 
%The key difference is that g-RISB introduces additional ghost orbitals to both the embedding impurity model and the quasiparticle Hamiltonian, allowing for systematic improvement in accuracy by increasing the number of ghost degrees of freedom ~\citep{gRISB_accuracy,Guerci_TDgGA,Guerci_thesis}.
%{\color{black}
The key concept of g-RISB is to enlarge the variational space of the original RISB by introducing auxiliary ``ghost'' degrees of freedom, where similar ideas have been developed simultaneously in different contexts of many-body approaches~\citep{Fertitta2018,Sriluckshmy2021,ancilla_quibits_2020,hidden_fermion}.
%}
It has been shown that g-RISB with two additional ghost orbitals provides a reliable description of the Mott transition in single-orbital Hubbard and Anderson lattice models when compared to DMFT ~\citep{gRISB_2017,gRISB_2021,Guerci2019}. Moreover, the studies on the one-band Hubbard model have shown that the accuracy of g-RISB can be systematically improved toward the exact solution in infinite dimensions by increasing the number of ghost orbitals~\citep{gRISB_accuracy,Guerci_thesis,Guerci_TDgGA}.
However, the accuracy of g-RISB as the number of ghost orbitals increases has not been explored in multiorbital systems.  Recently, the g-RISB has been applied to multiorbital systems focusing on the quasiparticle, spectral, and local atomic properties for multiorbital models on the Bethe lattice~\citep{Carlos2023}.
However, the energetics and the application of g-RISB to more realistic models of materials have not been explored. %{\color{black} %Moreover, it is interesting to explore if g-RISB can improve the aforementioned limitation of RISB, i.e., the overestimation of the critical Coulomb interaction of the Mott transition and the underestimation of the effective mass of Sr$_2$RuO$_4$~\cite{Lanata2019_NPJ,Facio_2018_PRB,Lechermann_2019_SRO}.
In particular, it would be desirable to explore if g-RISB can solve the above-mentioned limitations of RISB, i.e., the overestimation of the critical Coulomb interaction of the Mott transitions and the underestimation of the effective mass of Sr$_2$RuO$_4$~\cite{Lanata2019_NPJ,Facio_2018_PRB,Lechermann_2019_SRO}.
%}

In this work, we assess the accuracy of g-RISB on the degenerate three-orbital Hubbard model and the realistic Sr$_2$RuO$_4$ model extracted from first principle simulations, by comparing the results with DMFT. 
%{\color{black} We utilize the density matrix renormalization group (DMRG) method as a g-RISB impurity solver to investigate these models, allowing us to study the accuracy of g-RISB at a larger number of ghost orbitals beyond the capabilities of the exact-diagonalization (ED) solver.}
%The density matrix renoramlization group (DMRG) solver ~\cite{White1992,White1993} is utilized to assess the accuracy of g-RISB with larger bath orbitals in the embedded impurity model. 
For the degenerate three-orbital model, we demonstrate that g-RISB significantly enhances the accuracy of the original RISB approach, providing a more precise description of Hund's metallic behavior and Mott transition, where the critical point of the Mott transition is consistent with DMFT.
Furthermore, we provide numerical evidence that the accuracy of g-RISB can be systematically improved toward the exact DMFT limit in infinite dimensions by increasing the number of ghost orbitals in the considered multiorbital models. For
the realistic Sr$_2$RuO$_4$ model, our results illustrate that g-RISB captures reliable energy, quasiparticle weights, Fermi surface, and {\color{black}low-energy} spectral function in good agreement with DMFT and experiments.
%{\color{black} On the other hand, the high-energy Hubbard incoherent spectral features are approximated by the coherent bands due to the limitation of g-RISB that cannot capture a finite scattering rate with a small number of ghost orbitals.}
%Furthermore, we observe that the g-RISB energy is systematically more accurate than DMFT-ED in the considered electron fillings and Coulomb interactions. On the other hand, the g-RISB spectral functions and quasiparticle weight are also systematically improvable but converge slightly slower than DMFT-ED to the exact solution with increasing bath size. 
{\color{black} Moreover, we present the capability of employing the density matrix renormalization group (DMRG) method as the impurity solver within the g-RISB framework~\cite{White1992,White1993,Schollwock2005}, enabling us to explore systems characterized by a larger number of ghost orbitals that are beyond the capacity of the exact diagonalization (ED) solver.}
In addition, we discuss the connection between the g-RISB and DMFT self-energy.
Our benchmarks highlight the potential of g-RISB as a reliable tool for simulating strongly correlated materials.

\section{Model}

We consider the following three-orbital Hubbard model 

\begin{align}
H & =\sum_{\mathbf{k}}\sum_{ll'\sigma}\epsilon_{\mathbf{k}ll'}c_{\mathbf{k}l\sigma}^{\dagger}c_{\mathbf{k}l'\sigma}+\sum_{\mathbf{R}}H_{\mathbf{R}}^{\text{loc}}\big[\{c_{\mathbf{R}l\sigma}^{\dagger},c_{\mathbf{R}l\sigma}\}\big],\label{eq:H}
\end{align}
where $c^\dagger_{\mathbf{k}l\sigma}$ and $c_{\mathbf{k}l\sigma}$ are the electronic creation and the annihilation operators, respectively, with the orbital index
$l$, the spin index $\sigma$, the momentum index $\mathbf{k}$, and the site index $\mathbf{R}$, and $\epsilon_{\mathbf{k}ll'}$ is the hopping dispersion. In this work, we will focus on the $\epsilon_{\mathbf{k}ll'}$ corresponding to a degenerate semicircular density of states of the infinite-dimensional Bethe lattice and a realistic Sr$_2$RuO$_4$ dispersion extracted from the first principle simulations. The local Hamiltonian $H_{\mathbf{R}}^{\text{loc}}$ has the following form:
\begin{align}
H_{\mathbf{R}}^{\text{loc}}\big[\{c_{\mathbf{R}l\sigma}^{\dagger},c_{\mathbf{R}l\sigma}\}\big]=\sum_{ll'\sigma}&\epsilon^{\text{loc}}_{\mathbf{R},ll'}c_{\mathbf{R}l\sigma}^{\dagger}c_{\mathbf{R}l'\sigma}\nonumber\\
&+H_{\mathbf{R}}^{\text{int}}\big[\{c_{\mathbf{R}l\sigma}^{\dagger},c_{\mathbf{R}l\sigma}\}\big],
\end{align}
where $\epsilon_{\mathbf{R}ll'}^\text{loc}$ is the local one-body interaction and 
\begin{align}
H_{\mathbf{R}}^{\text{int}} & \big[\{c_{\mathbf{R}l\sigma}^{\dagger},c_{\mathbf{R}l\sigma}\}\big]=U\sum_{l}n_{\mathbf{R}l\uparrow}n_{\mathbf{R}l\downarrow}\nonumber\\
&+U'\sum_{l<l',\sigma}n_{\mathbf{R}l\sigma}n_{\mathbf{R}l'\bar{\sigma}} +(U'-J)\sum_{l<l',\sigma}n_{\mathbf{R}l\sigma}n_{\mathbf{R}l'\sigma}\nonumber \\
 &-J\sum_{l<l'}\big(c_{\mathbf{R}l\uparrow}^{\dagger}c_{\mathbf{R}l\downarrow}c_{\mathbf{R}l'\downarrow}^{\dagger}c_{\mathbf{R}l'\uparrow}+c_{\mathbf{R}l\uparrow}^{\dagger}c_{\mathbf{R}l\downarrow}^{\dagger}c_{\mathbf{R}l'\uparrow}c_{\mathbf{R}l'\downarrow}\nonumber\\
 &+\text{H.c.}\big).\label{eq:Hint}
\end{align}
is  the Hubbard-Kanamori parameterization of the two-body
Coulomb interaction~\cite{Kanamori_1963}. In addition, we use the rotationally-invariant condition $U'=U-2J$. In the following, we will also introduce the index $\alpha\equiv(l,\sigma)$
to combine the electronic spin and orbital degrees of freedom.

\section{Methods}

\subsection{Ghost rotationally-invariant slave-boson}

Here, we describe the formalism of g-RISB for multiorbital models, which is the direct generalization of the previous works \citep{gRISB_2017,gRISB_2021,gRISB_2022,gRISB_accuracy}. The g-RISB theory is encoded in the following Lagrange function:

\begin{align}
\mathcal{L}[\Phi, & E^{c};R,\lambda;D,\lambda^{c};\Delta,\Psi_{0},E]=\frac{1}{\mathcal{N}}\langle\Psi_{0}|\hat{H}^{\text{qp}}[R,\lambda]|\Psi_{0}\rangle\nonumber \\
 & +E(1-\langle\Psi_{0}|\Psi_{0}\rangle)+\Big[\langle\Phi|\hat{H}^{\text{emb}}[D,\lambda^{c}]|\Phi\rangle\nonumber \\
 & +E^{c}(1-\langle\Phi|\Phi\rangle)\Big]-\Big[\sum_{ab}\big[\lambda+\lambda^{c}\big]_{ab}\big[\Delta\big]_{ab}\nonumber \\
 & +\sum_{ca\alpha}\big(\big[D\big]_{a\alpha}\big[R\big]_{c\alpha}\big[\Delta(1-\Delta)\big]_{ca}^{\frac{1}{2}}+\text{c.c.}\big)\Big],
\end{align}
where $H^{\text{qp}}$ is the quasiparticle Hamiltonian, $H^{\text{emb}}$ is the embedding Hamiltonian, and $|\Psi_{0}\rangle$ and $|\Phi_{i}\rangle$ are their corresponding wavefunction, respectively. The $E$ and $E^c$ are Lagrange multipliers enforcing the normalization of the wavefunctions. The $\lambda$ and $\lambda^c$ enforce the structure of the quasiparticle density matrix $\Delta$, and $D$ enforces the structure of the quasiparticle renormalization matrix $R$. The $\mathcal{N}$ is the total number of sites in the lattice.

\begin{figure}[t]
\begin{centering}
\includegraphics[scale=0.36]{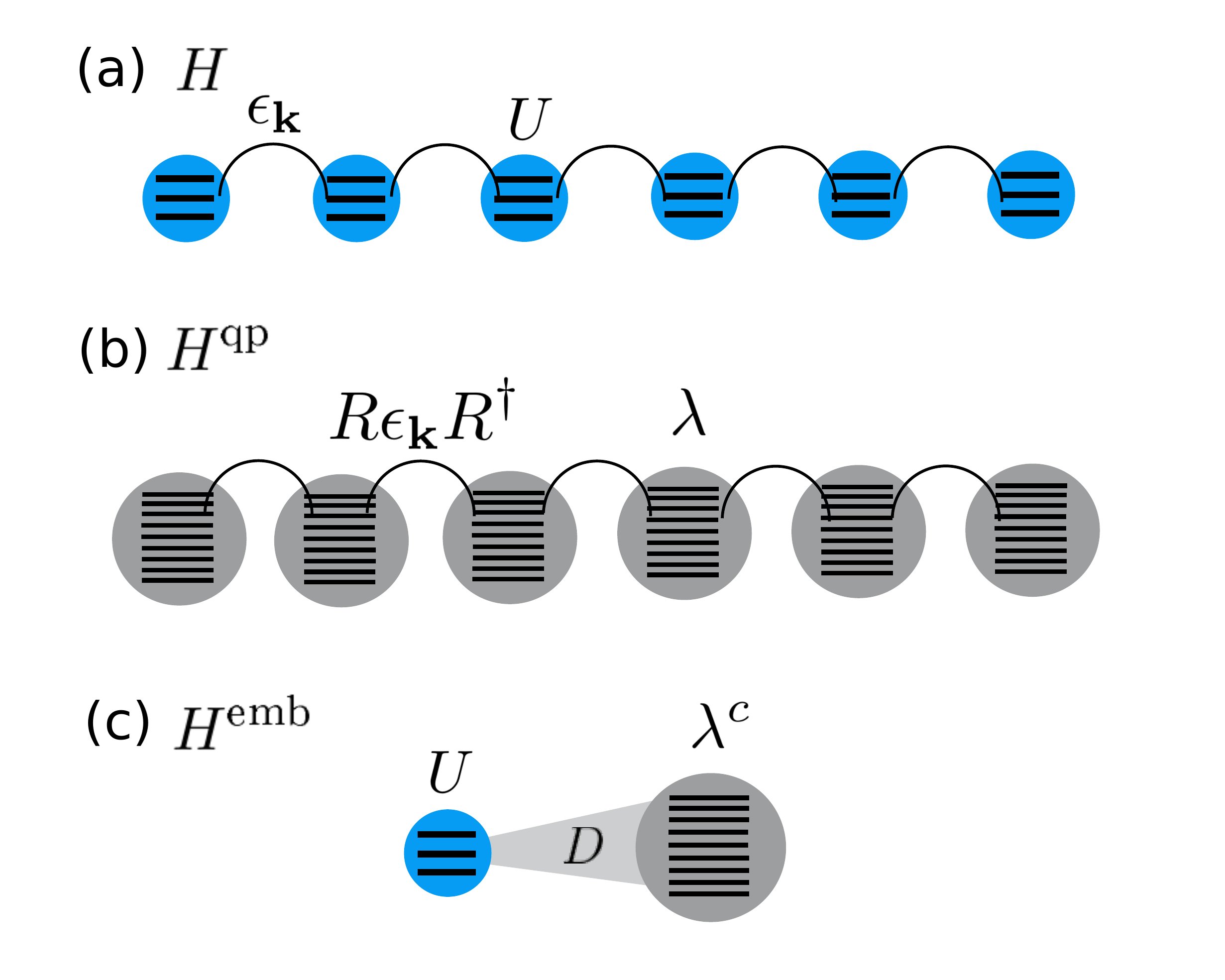}
\par\end{centering}
\caption{Schematic representation of the (a) original three-orbital Hubbard Hamiltonian $H$, (b) the non-interacting quasiparticle Hamiltonian $H^{\text{qp}}$ including enlarged auxiliary ghost degrees of freedom with in total nine orbitals, and (c) the embedding Hamiltonian $H^{\text{emb}}$ including an impurity with three orbitals and a bath with nine orbitals. The number of the orbitals in the bath of $H^{\text{emb}}$ and in $H^{\text{qp}}$ can be increased simultaneously by introducing more ghost orbitals to improve the g-RISB accuracy. Here we illustrate the Hamiltonians at bath size $N_b=9$. \label{fig:scheme}}
\end{figure}

The quasiparticle Hamiltonian has the following form:
\begin{equation}
H^{\text{qp}}=\sum_{\mathbf{k}}\sum_{ab}\sum_{\alpha\beta}\Big[R_{a\alpha}\epsilon_{\mathbf{k},\alpha\beta}R_{\beta b}^{\dagger}+\lambda_{ab}\Big]f_{\mathbf{k}a}^{\dagger}f_{\mathbf{k}b},\label{eq:Hqp}
\end{equation}
where, for the three-orbital model, $\alpha,\beta\in\{1\uparrow,1\downarrow,2\uparrow,2\downarrow,3\uparrow,3\downarrow\}$ labels the original physical degrees of freedom $c_{\mathbf{k}\alpha}$, and $a,b$ labels the auxiliary quasiparticle degrees of freedom $f_{\mathbf{k} a}$. %{\color{black}The physical and the quasiparticle operators are related through $c_{\mathbf{k}\alpha}^\dagger= R_{a\alpha} f_{\mathbf{k}a}^\dagger$ at the mean-field level~\cite{Lecherman_2007,Lanata_2017_PRL,gRISB_2022}.}
The number of quasiparticle orbitals can be systematically increased by adding the so-called ghost orbitals to improve the accuracy of g-RISB. In this work, we used up to fifteen auxiliary quasiparticle orbitals, i.e., $a,b\in\{1\uparrow,1\downarrow,....,15\uparrow,15\downarrow\}$ to study the convergence behavior of g-RISB in multiorbital Hubbard models. Note that with the minimal three quasiparticle orbitals, $a,b\in\{1\uparrow,1\downarrow,2\uparrow,2,\downarrow3\uparrow,3\downarrow\}$, g-RISB recovers the original RISB approach.

The embedding Hamiltonian is as follows:

\begin{align}
\hat{H}^{\text{emb}} & =\hat{H}^{\text{loc}}\big[\{\hat{c}_{\alpha}^{\dagger},\hat{c}_{\alpha}\}\big]+\sum_{a\alpha}\Big(D_{a\alpha}\hat{c}_{\alpha}^{\dagger}\hat{f}_{a}\nonumber \\
 & +\text{H.c.}\Big)+\sum_{ab}\lambda_{ab}^{c}\hat{f}_{b}\hat{f}_{a}^{\dagger},\label{eq:Hemb}
\end{align}
where $\hat{c}^\dagger_{\alpha}$ and  $\hat{c}_{\alpha}$ are the impurity creation and annihilation operators, respectively, and $\hat{f}^\dagger_{a}$ and $\hat{f}_{a}$ are the bath creation and annihilation operators, respectively. We will use $N_b$ to label the number of the bath orbitals. Note that the number of the bath orbitals is the same as the number of the quasiparticle orbitals in $H^{\text{qp}}$.
%, which we will use up to $N_b=15$ orbitals in this work. 
%The g-RISB embedding Hamiltonian has a similar structure as the quantum impurity model in DMFT when a finite number of bath orbitals are used to represent the hybridization function~\citep{Caffarel_Krauth_1994,Rozenberg1994}.
%(As in g-RISB, the accuracy of DMFT can be systematically improved by increasing the number of bath orbitals.) 
The schematic representation of the two models is shown in Fig. \ref{fig:scheme}.

The stationary condition of the g-RISB Lagrange function leads to the following saddle-point equations:

\begin{gather}
\frac{1}{\mathcal{N}}\Big[\sum_{\mathbf{k}}n_{F}(R\epsilon_{\mathbf{k}}R^{\dagger}+\lambda)\Big]_{ba}=\Delta_{ab}\label{eq:SP1}\\
\frac{1}{\mathcal{N}}\Big[\sum_{\mathbf{k}}\epsilon_{\mathbf{k}}R^{\dagger}n_{F}(R\epsilon_{\mathbf{k}}R^{\dagger}+\lambda)\Big]_{ba}=\sum_{ac\alpha}D_{c\alpha}\big[\Delta(I-\Delta)\big]_{ac}^{\frac{1}{2}}\label{eq:SP2}\\
\sum_{cd\alpha}\frac{\partial}{\partial \Delta_{ab}}\Big(\big[\Delta(I-\Delta)\big]_{cd}^{\frac{1}{2}}D_{d\alpha}R_{c\alpha}+\text{c.c.}\Big)+\big[\lambda+\lambda^{c}]_{ab}=0\label{eq:SP3}\\
\hat{H}^{\text{emb}}|\Phi\rangle=E^{c}|\Phi\rangle\label{eq:SP4}\\
\langle\Phi|\hat{c}_{\alpha}^{\dagger}\hat{f}_{a}|\Phi\rangle-\sum_{c}\big[\Delta(I-\Delta)\big]_{ac}^{\frac{1}{2}}R_{c\alpha}=0\label{eq:SP5}\\
\langle\Phi|\hat{f}_{b}\hat{f}_{a}^{\dagger}|\Phi\rangle-\Delta_{ab}=0\label{eq:SP6}
\end{gather}
where $n_F$ is the Fermi function, $I$ is the identity matrix, and the variables $R$, $\lambda$, $\Delta$, $D$, $\lambda^{c}$, and $|\Phi\rangle$ are determined self-consistently. 
%Note that Eq.~\ref{eq:SP1}-\ref{eq:SP6} corresponds to the saddle-point approximation to the g-RISB functional (see Refs.~\onlinecite{gRISB_2017,gRISB_2022}).%, which cannot capture the incoherent features in the spectral functions and a finite scattering rate. 
%Therefore, all the bands in the g-RISB spectral functions are described by coherent bands, including the Hubbard bands, as we will see in the next section.
{\color{black}Equations~\ref{eq:SP1}-\ref{eq:SP6} have been rigorously derived in Refs.~\onlinecite{gRISB_2017,gRISB_2021,gRISB_2022}. Here, we provide a simplified physical rationale to elucidate their meaning following Refs.~\onlinecite{Lanata_2015_PRX,RISB_DMET_Ayral_2018,gDMET}.
As in any quantum embedding method, there are three basic ingredients: a high-level description of a local system (the embedding Hamiltonian $\hat{H}^\text{emb}$ parameterized by $D$ and $\lambda^c$), a low-level description of the lattice system (the quasiparticle Hamiltonian $H^{\text{qp}}$ parameterized by $R$ and $\lambda$), and compatibility conditions that establish a self-consistent linkage between the parameters of these two systems. 
For a given set of parameters, equations~\ref{eq:SP1} and~\ref{eq:SP4} provide us with the density matrix and the ground state wavefunction of the two reference systems, respectively. The remaining saddle-point equations then utilize these quantities to determine a new set of parameters of both systems for self-consistency calculations (see Appx.~\ref{sec:algorithm} for our algorithm).
%The parameter $D$ and $\lambda^c$ (see Eq.~\ref{eq:Hemb}) parameterize the quantum impurity model and its ground state (Eq.~\ref{eq:SP4}).
%The parameters $R$ and $\lambda$ parameterize the ``quasiparticle'' band structure constituting the bath via the quasiparticle Hamiltonian (Eq.~\ref{eq:Hqp}). 
%Equation~\ref{eq:SP1} states that $\Delta$ is the quasiparticle density matrix, and Eq.~\ref{eq:SP2} expresses $D$ in terms of a quantity related to the quasiparticle kinetic energy. 
%Equation~\ref{eq:SP6} equates the density matrix of the impurity bath to the local density matrix of the low-level description quasiparticles density matrix $\Delta$.  
%Equation~\ref{eq:SP5} states how the local version of the kinetic energy, namely the hybridization of the fragment and its bath is related to the quasiparticle renormalization parametrized by $R$.
%Equation~\ref{eq:SP3}, completes the full set of self-consistency equations by relating the difference between the shifts of the quasiparticle energies and bath electron energies to the kinetic energy. 
}

{\color{black}
For the impurity solver of $H_{\text{emb}}$, we utilize the Lanczos ED algorithm to compute the ground state wavefunction $|\Phi\rangle$ for bath size $N_{b}\leq9$. To explore the accuracy at a larger bath size $N_{b}=15$, it is necessary to utilize the DMRG approach to solve the ground state wavefunction~\cite{White1992,White1993,Schollwock2005}, as the Hilbert space is too large for performing the Lanczos ED. In this work, we utilize the DMRG implemented in Block2 and PySCF as our impurity solver \citep{PySCF_2018,PySCF_2020,block2,block2_2}.
}

With the converged $R$ and $\lambda$, one can compute the Green's function from 
\begin{equation}
G_{\alpha\beta}(\mathbf{k},\omega)=\sum_{ab}R_{\alpha a}^{\dagger}\big[(\omega+i0^{+})I-R\epsilon_{\mathbf{k}}R^{\dagger}-\lambda\big]_{ab}^{-1}R_{b\beta}\label{eq:G}
\end{equation}
and the self-energy can be determined from the Dyson equation
\begin{equation}
\Sigma_{\alpha\beta}(\omega)=\big[G_{0}^{-1}(\mathbf{k},\omega)-G^{-1}(\mathbf{k},\omega)\big]_{\alpha\beta},
\end{equation}
where $G_0(\mathbf{k},\omega)$ is the bare Green's function. The quasiparticle renormalization weight is determined from 
\begin{equation}
Z_{\alpha\beta}=\left[1-\frac{\partial\text{Re}\Sigma(\omega)}{\partial\omega}\Big|_{\omega\rightarrow0}\right]^{-1}_{\alpha\beta}.
\end{equation}

%{\color{black}
An expression for the g-RISB  self-energy has been derived in the previous work%exploiting the gauge degrees of freedom in g-RISB
~\citep{gRISB_2017,gRISB_2021,gDMET}, resulting into a pole-expansion representation. 
In this work,  we exploit  the gauge  freedom
\citep{Lecherman_2007,gRISB_2017,gRISB_2021}
to provide a more compact expression, and provide further insight into the  connection between RISB and   the  DMFT self-energy. Specifically, we choose a gauge transforming $R$ and $\lambda$ into the following block matrix form:
\begin{align}
u^\dagger R=\tilde{R}=\begin{pmatrix}\tilde{R}_0\\
0
\end{pmatrix},\ 
u^\dagger \lambda u=\tilde{\lambda}=\begin{pmatrix}\tilde{\lambda}_0 & \tilde{\lambda}_1\\
\tilde{\lambda}_1^\dagger & \tilde{\lambda}_{2}
\end{pmatrix},\label{eq:gauge}
\end{align}
where $u$ is the gauge transformation matrix~\citep{Lecherman_2007,Lanata_2017_PRL}, $\tilde{R}^0$ and $\tilde{\lambda}^0$ are square matrices of size $N\times N$, where $N$ is the total number of physical spin orbitals. The matrix $\tilde{\lambda}_1$ is a $N\times M$ matrix where $M=2N_b-N$, and $\tilde{\lambda}_2$ is a $M\times M$ diagonal matrix. {\color{black} This specific gauge always exist and the transformation matrix $u$ can be calculated using the singular value decomposition method~\cite{gDMET}.}
From Eq.~\ref{eq:G}, one can show that the self-energy has the following form:
\begin{align}
\Sigma_{\alpha\beta}&(\omega)=\omega\big[I-\big[\tilde{R}_0^{\dagger}\tilde{R}_0\big]^{-1}\big]_{\alpha\beta}+\sum_{ab}\big[\tilde{R}_0\big]^{-1}_{\alpha a}\big[\tilde{\lambda}_0\big]_{ab}\big[\tilde{R}^{\dagger}_0\big]_{b\beta}^{-1}\nonumber\\
&-\big[\epsilon^\text{loc}\big]_{\alpha\beta}+\sum_{abc}\big[\tilde{R}_0\big]_{\alpha a}^{-1}\frac{\big[\tilde{\lambda}_1\big]_{ac}\big[\tilde{\lambda}_1^\dagger\big]_{bc}}{\big[(\omega+i0^+)I-\tilde{\lambda}_2\big]_{cc}}\big[R^{\dagger}_0\big]_{b\beta}^{-1} .\label{eq:sig}
\end{align}
In this work, we provide numerical evidence that $\tilde{R}^\dagger \tilde{R}$, {\color{black} which equals the total spectral weight:
\begin{equation}
\big[\tilde{R}^\dagger \tilde{R}\big]_{\alpha\beta}=-\frac{1}{\mathcal{N}\pi} \sum_{\mathbf{k}}\int_{-\infty}^{\infty}d\omega \text{Im} \big[G(\mathbf{k},\omega)\big]_{\alpha\beta},
\end{equation}}approaches the identity as one increases the number of the ghost orbitals (see Appx.~\ref{sec:RdR_Siginf}). In this limit, the expression above simplifies further, as the first term in Eq. 17 vanishes.  We have also checked  numerically that 
$\Sigma(\infty)$ approaches the Hartree-Fock self-energy with increasing ghost orbitals (see Appx.~\ref{sec:RdR_Siginf}). The equation above, obtained from g-RISB principles, closely resembles the pole expansion proposed in previous DMFT literature~\citep{Savrasov2006}.%, helping to formalize the connection between the self-energy of g-RISB and DMFT. 
%}
{\color{black} Also, Eq. ~\ref{eq:sig} implies the number of poles in the self-energy grows linearly with the number of ghost orbitals.}

\subsection{Dynamical mean-field theory}

We also perform the DMFT calculations to compare with g-RISB. To benchmark the accuracy with respect to the number of the bath orbitals, we utilize DMFT with the Lanczos ED solver (DMFT-ED) with discretized bath orbitals, which has been applied to multiorbital models \citep{Caffarel_Krauth_1994,Rozenberg1994,Liebsch_2005,Liebsch_2007,Liebsch_2010,Liebsch_2011,Capone2004,Civelli_2005,Kyung_2006,AMARICCI_EDI,ISKAKOV2018128}. The hybridization function scheme is used for the bath fitting procedure with weight function $1/\omega_{n}$ \citep{Capone2007}, allowing stabler convergences for all the considered parameters. A fictitious temperature $\beta=200$ is introduced for the bath fitting. The Lanczos algorithm is utilized to compute the ground state wavefunctions and Green's functions. In addition, we performed the DMFT calculations with the continuous-time quantum Monte Carlo (CTQMC) solver implemented in the TRIQS library, providing us the exact solution for the benchmark \citep{triqs,TRIQS_CTHYB,Gull_RMP_2011,MELNICK2021108075}.

\begin{figure}[t]
\begin{centering}
\includegraphics[scale=0.4]{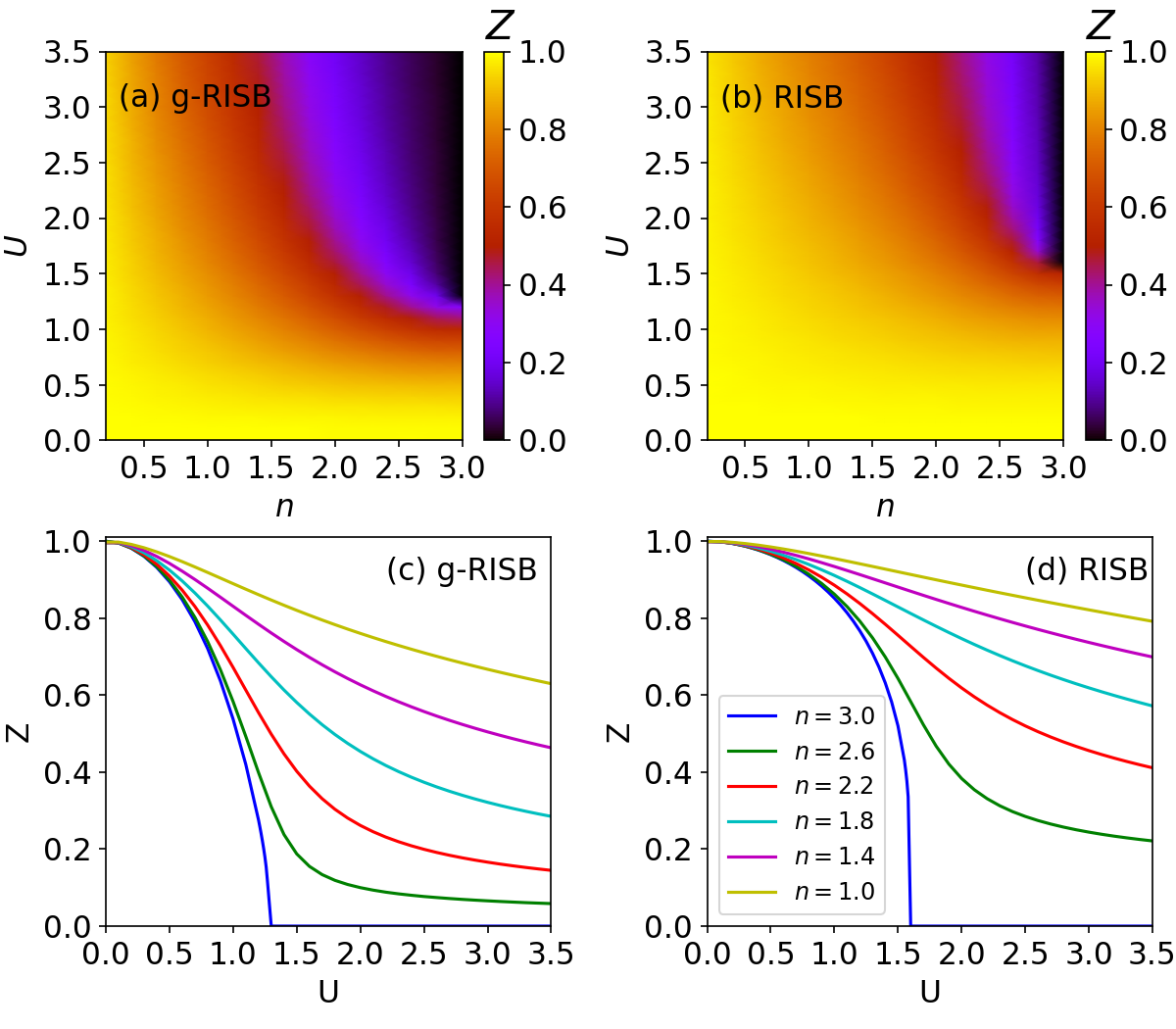}
\par\end{centering}
\caption{(a) and (c) the quasiparticle weights $Z$ for the degenerate three-orbital Hubbard
model on Bethe lattice from g-RISB with $N_b=9$ bath orbitals, and (b) and (d) the quasiparticle weights of RISB (g-RISB with $N_b=3$) as a function of
Coulomb interaction $U$ with $J=0.25U$ for different electron filling $n$. {\color{black}The energy unit is the half-bandwidth.} \label{fig:Z_deg}}
\end{figure}

\section{Results}

\subsection{Degenerate three-orbital Hubbard model}

We first apply the g-RISB approach to the degenerate three-orbital Hubbard model focusing on Hund's metal regime with Hund's coupling fixed at $J=0.25U$, where the positive Hund's rule coupling generates high-spin incoherent Hund's metallic states away from half-filling \citep{Georges_Hunds_review,Medici_Janus_PRL,HauleHunds}. We consider the semicircular density of states $N(\omega)=\frac{1}{\pi t}\sqrt{1-[\omega/(2 t)]^2}$ with the energy unit set to half-bandwidth $D=2t=1$, corresponding to the $\epsilon_{\mathbf{k}l\sigma}$ on the Bethe lattice in the limit of infinite coordination \citep{DMFT_RMP_1996}.

The quasiparticle weights $Z$ for g-RISB and RISB are shown in Fig. \ref{fig:Z_deg}. From the density plot of the quasiparticle weights in Fig. \ref{fig:Z_deg}(a) and (b), we found that the 
%Hund's metallic regime, characterized by small quasiparticle weights (the purple regime), 
{\color{black} g-RISB quasiparticle weight is much smaller than the original RISB approach in the Hund's metal regime.} In particular, for the electron filling $n=2$ relevant for Sr$_{2}$RuO$_{4}$ and iron-based superconductors, the quasiparticle weights are around $Z\gtrsim 0.5$ even with large Coulomb interactions. On the other hand, the quasiparticle weights in g-RISB are about $Z\sim0.2$ at large Coulomb interactions, which are strongly correlated. In Fig. \ref{fig:Z_deg}(c) and (d), we show the corresponding quasiparticle weights $Z$ as a function of Coulomb interaction for various fillings $n$. At half-filling $n=3$, the Mott transition for g-RISB and RISB is at $U_{c}=1.25$ and $U_{c}=1.6$, respectively. We found that RISB is quantitatively reliable only around half-filling $n=3$ and becomes less accurate away from half-filling. The overestimation of the quasiparticle weights in RISB, i.e., the underestimation of the effective mass, is also observed in the previous studies of Sr$_{2}$RuO$_{4}$, whose values deviate significantly from the DMFT values and experiments even using larger Coulomb parameters \citep{Facio_2018_PRB,Lechermann_2019_SRO}.

\begin{figure}[t]
\begin{centering}
\includegraphics[scale=0.4]{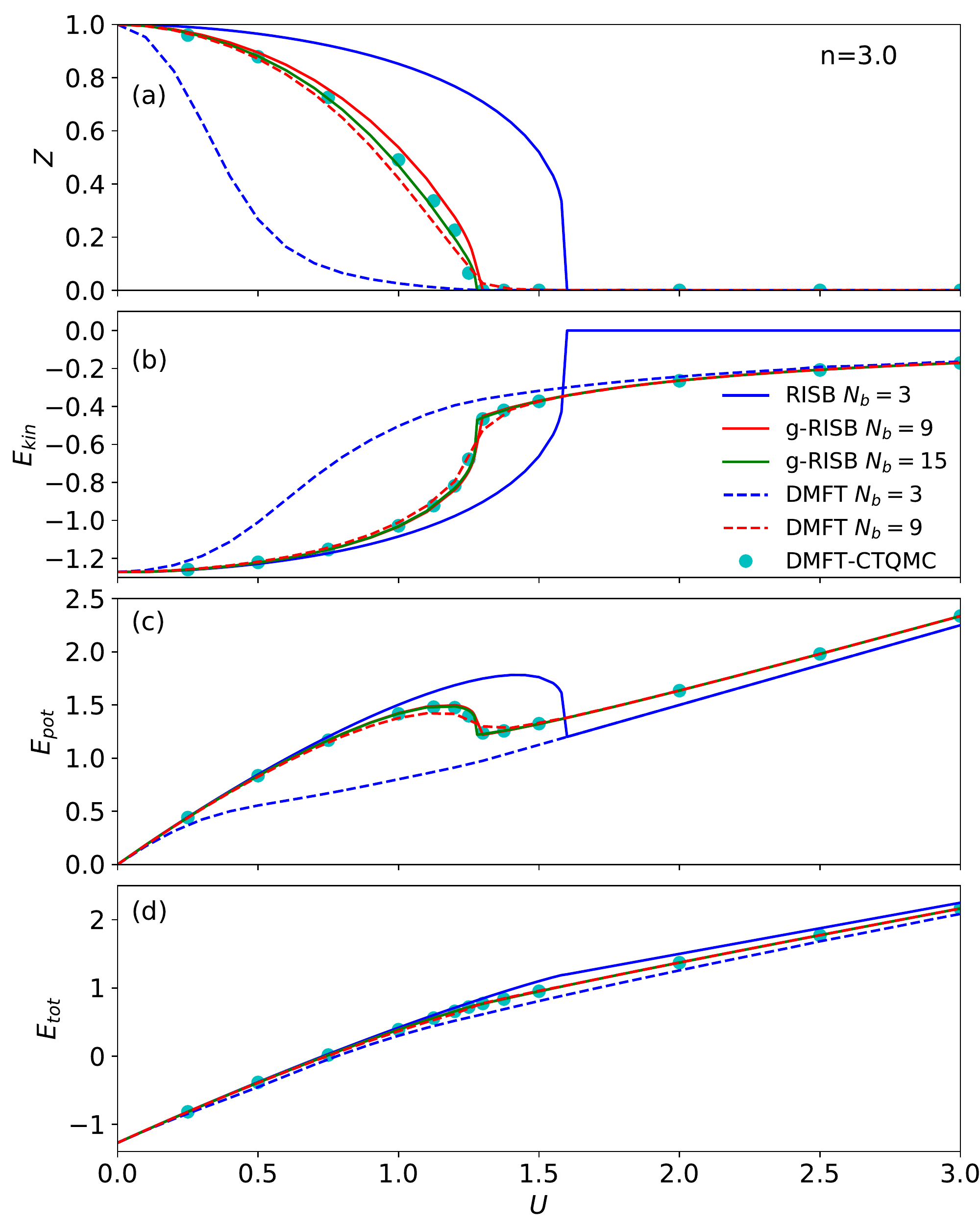}
\par\end{centering}
\caption{(a) The quasiparticle weights $Z$, (b) kinetic energy $E_{\text{kin}}$, (c) potential energy $E_{\text{pot}}$, and (d) total energy $E_{\text{tot}}$ for the degenerate three-orbital model on Bethe lattice for g-RISB with bath size $N_{b}=3,\ 9,\ 15$ and DMFT-ED with $N_{b}=3,\ 9$ as a function of Coulomb interaction $U$ with Hund's coupling $J=0.25U$ at filling $n=3$. The DMFT-CTQMC results at inverse temperature $\beta=200$ are shown for comparison. {\color{black}The energy unit is the half-bandwidth.} \label{fig:benchmark_n3}}
\end{figure}

\begin{figure}[t]
\begin{centering}
\includegraphics[scale=0.4]{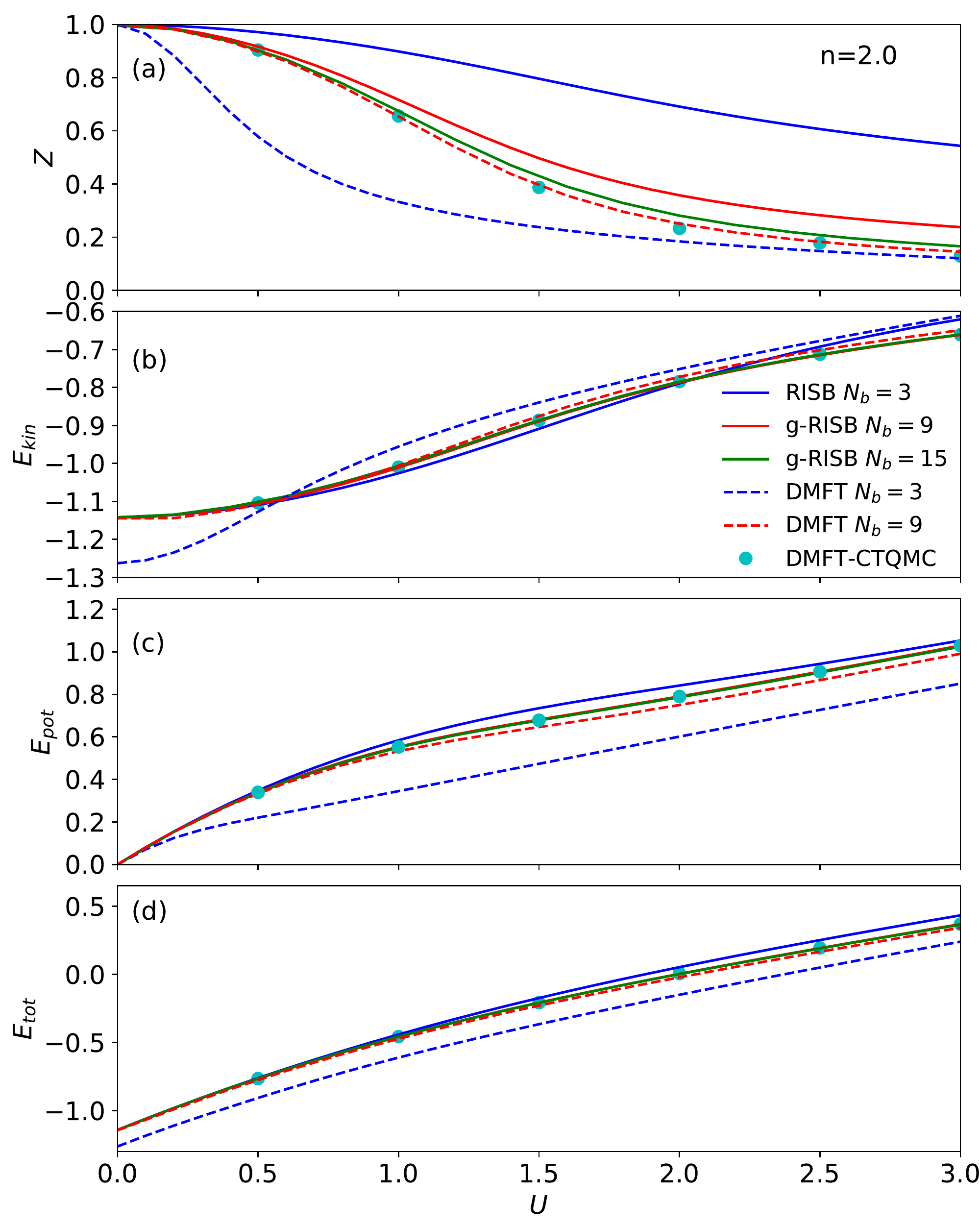}
\par\end{centering}
\caption{(a) The quasiparticle weights $Z$, (b) kinetic energy $E_{\text{kin}}$, (c) potential energy $E_{\text{pot}}$, and (d) total energy $E_{\text{tot}}$ for the degenerate three-orbital model on Bethe lattice for g-RISB with bath size $N_{b}=3,\ 9,\ 15$ and DMFT-ED with $N_{b}=3,\ 9$ as a function of Coulomb interaction $U$ with Hund's coupling $J=0.25U$ at filling $n=2$. The DMFT-CTQMC results at inverse temperature $\beta=200$ are shown for comparison. {\color{black}The energy unit is the half-bandwidth.} \label{fig:benchmark_n2}}
\end{figure}

We now examine the accuracy of g-RISB closely as a function of Coulomb interactions at half-filling $n=3$. The corresponding quasiparticle weights $Z$, kinetic energy $E^{\text{kin}}$, potential energy $E^{\text{pot}}$, and total energy $E^{\text{tot}}$ are shown in Fig. \ref{fig:benchmark_n3} for g-RISB, DMFT-ED, and DMFT-CTQMC. We observed that the g-RISB energy and quasiparticle weights converge systematically to the DMFT-CTQMC values with increasing bath orbitals $N_b$. With bath size $N_{b}=9$, g-RISB already gives reliable energy and quasiparticle weights close to the exact DMFT-CTQMC values.   On the other hand, the DMFT-ED results still have a slight discrepancy from the exact DMFT-CTQMC values, especially around the metal-insulator transition $U_{c}$. Moreover, g-RISB captures more precisely the discontinuity in the quasiparticle weights and the energy at $U_{c}$~\citep{DeMedici2022}, and the accuracy in the quasiparticle weights can be further improved by increasing the bath size to $N_{b}=15$. To have a closer look at the accuracy of g-RISB, we also show the g-RISB total energy $E_{\text{tot}}$ for the selected Coulomb interactions $U$ for different bath sizes $N_b$ in Tab.~\ref{tab:energy_deg}.

In Fig.~\ref{fig:benchmark_n2}, we benchmarked the accuracy of g-RISB as a function of Coulomb interactions $U$ at electron filling $n=2$, corresponding to the parameter regime of Hund's metal materials, e.g., Sr$_{2}$RuO$_{4}$ and iron-based superconductors. We observed that the accuracy of g-RISB is systematically improvable with the increasing number of bath orbitals $N_{b}$. With $N_{b}=9$, g-RISB produces accurate energy close to the exact DMFT-CTQMC values, while DMFT-ED shows slight differences from the exact DMFT-CTQMC values. On the other hand, DMFT-ED gives accurate quasiparticle weights $Z$ at $N_{b}=9$, while g-RISB requires $N_{b}=15$ bath orbitals to have similar accuracy. %These results suggest that g-RISB is an accurate approach for studying the energy of correlated systems with small bath orbitals, while the spectral properties are slightly less accurate than DMFT, with the same number of bath orbitals. 
To have a closer look at the g-RISB energy, we show the total energy for the selected parameters in Tab.~\ref{tab:energy_deg}.

\begin{figure}[t]
\begin{centering}
\includegraphics[scale=0.41]{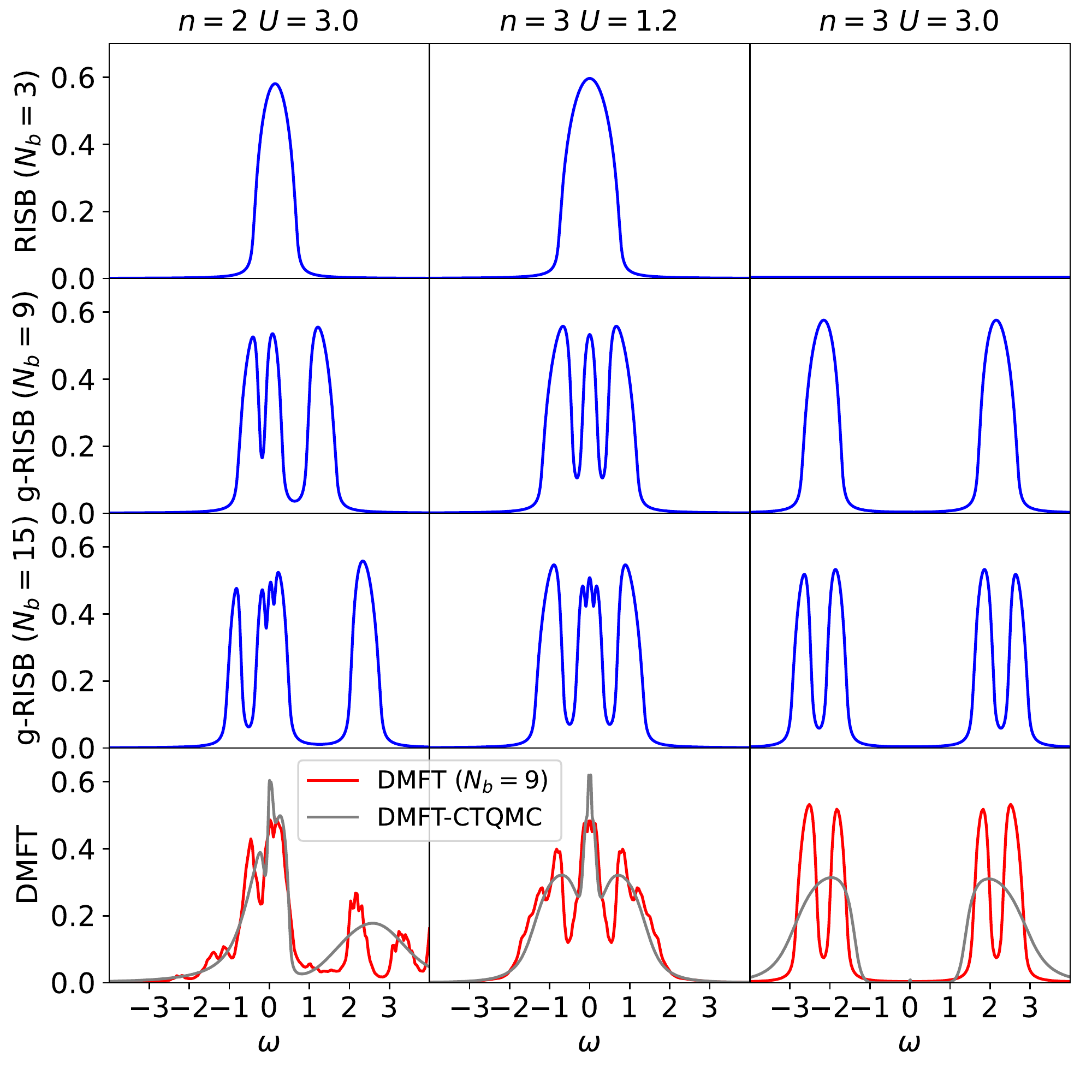}
\par\end{centering}
\caption{The g-RISB spectral functions $A(\omega)$ with bath size $N_{b}=3,\ 9,\ 15$ compared with the DMFT spectral function using ED solver with $N_{b}=9$ and CTQMC solver for electron filling $n=2$ and Coulomb $U=3$ (the first column), $n=3$ and $U=1.2$ (the second column), and $n=3$ and $U=3$ (the third column). The Hund's coupling is $J=0.25U$. The inverse temperature in the DMFT-CTQMC calculations is $\beta=200$, and {\color{black}the energy unit is the half-bandwidth.} {\color{black} A broadening factor $\eta=0.05$ is used in the ED and g-RISB.} \label{fig:dos_deg}}
\end{figure}

\begin{table}
\begin{tabular}{c c c c c c} 
 \hline
 n\;\;\; & U\;\;\;\; & $N_b=3$\;\; & $N_b=9$\;\; & $N_b=15$\;\;  & DMFT-CTQMC \\ [0.5ex]  
\hline\hline
3.0\;\; & 1.0\;\; & 0.412 &  0.389 & 0.388 & 0.389 \\ 
 \hline
3.0\;\; & 2.5\;\; & 1.875 & 1.774 & 1.773 & 1.772 \\
\hline\hline
2.0\;\; & 1.5\;\; & -0.174 & -0.208 & -0.209 & -0.208 \\ 
 \hline
2.0\;\; & 2.5\;\; & 0.2506 & 0.192 & 0.189 & 0.193 \\
 \hline
\end{tabular}
\caption{The g-RISB total energy $E_{\text{tot}}$ for the degenerate three-orbital model on Bethe lattice for the selected fillings $n$ and $U$ with different numbers of bath orbitals $N_b$ and $J=0.25U$. The DMFT energy at $\beta=200$ with CTQMC solver is shown for comparison. The energy unit is the half-bandwidth. \label{tab:energy_deg}} 
\end{table}

Figure \ref{fig:dos_deg} shows the spectral functions $A(\omega)=-\text{Im}G(\omega)/\pi$ for several sets of parameters in the degenerate three-orbital model. For RISB $N_{b}=3$, the metal-insulator transition is of the Brinkman-Rice scenario, where only the band renormalization around the Fermi level is captured, and the incoherent Hubbard bands are absent. Therefore, for $n=3$ and $U=3$, the spectral function is zero, which is a crude approximation to the Mott insulator. On the other hand, for $N_{b}=9$, g-RISB can capture both the coherent quasiparticle peak and the Hubbard bands, and the quality of the spectral functions can be systematically improved by adding more bath orbitals to $N_{b}=15$. Furthermore, the overall position of the peaks in the g-RISB spectral functions is in good agreement with the DMFT-ED and DMFT-CTQMC spectral functions. These results suggest that while g-RISB can capture reliable energy, the spectral functions require more bath orbitals to reach a similar accuracy to DMFT-ED. {\color{black} In addition, since g-RISB with a small number of ghost orbitals cannot reliably capture the finite scattering rate in the imaginary part of the self-energy (see Appx.~\ref{sec:self-energy}), its high-energy spectral function becomes more coherent than DMFT-CTQMC.} {\color{black} This is also a feature in DMFT-ED with a small number of bath orbitals, where the exponentially large number of poles tends to cluster together forming sharp features in the spectrum, see Fig.~\ref{fig:dos_deg} and Ref.~\onlinecite{Caffarel_Krauth_1994,Rozenberg1994,DMFT_RMP_1996}.}

\begin{figure}[t]
\begin{centering}
\includegraphics[scale=0.4]{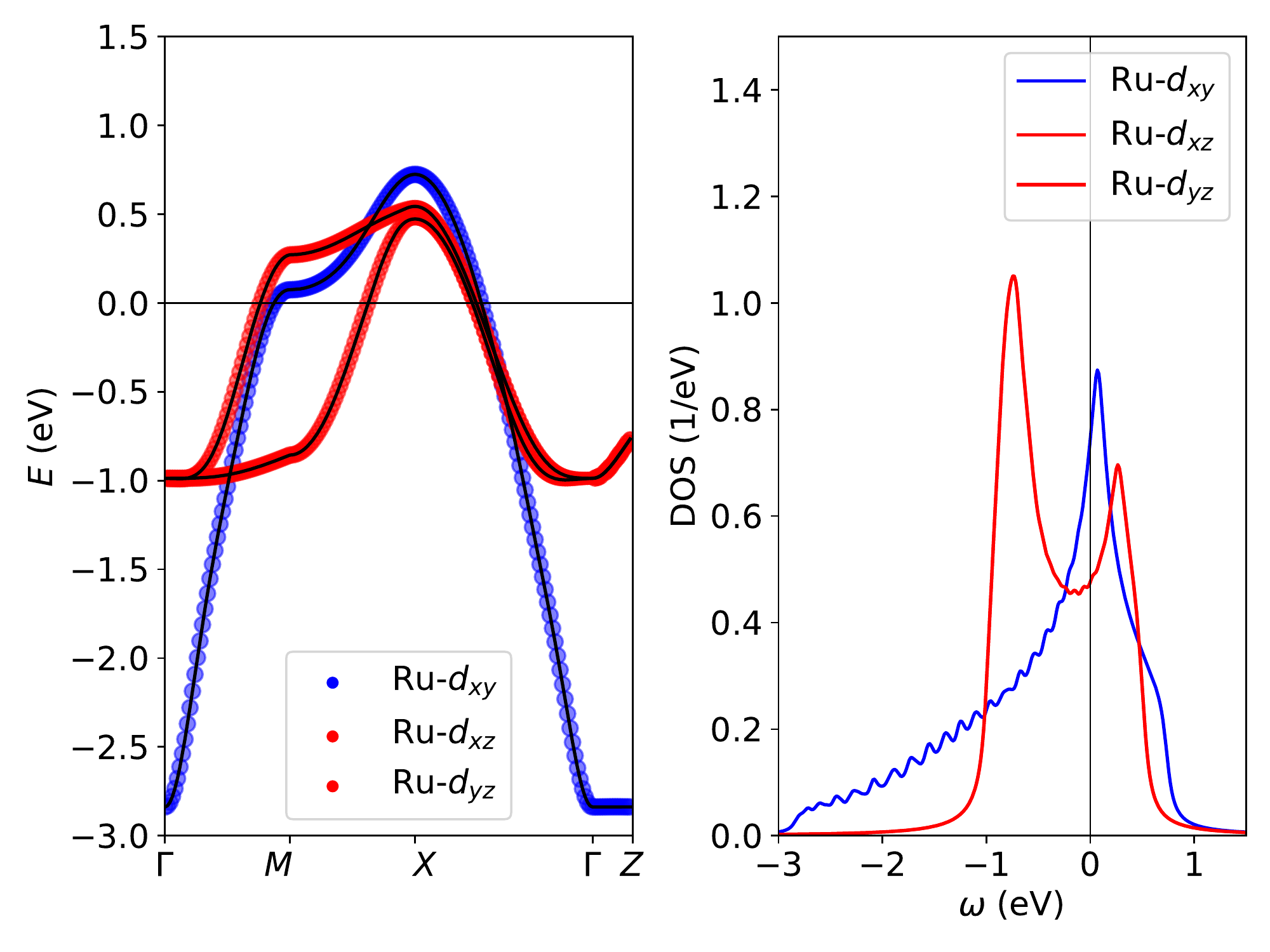}
\par\end{centering}
\caption{(a) The Wannier tight-binding bandstructure for Sr$_2$RuO$_4$ extracted from density functional theory along the high-symmetry path in the Brillouin Zone and (b) the orbital-resolved density of states constructed from the density functional theory considering the Ru-$d_{xy}$, $d_{xz}$, and $d_{yz}$ orbitals. {\color{black} A broadening factor $\eta=0.05$ is used in the density of states.} \label{fig:SRO_DFT}}
\end{figure}

\subsection{Realistic Sr$_{2}$RuO$_{4}$ model}

In this subsection, we present the results for Sr$_{2}$RuO$_{4}$, which is a well-characterized bad metal driven by Hund's physics \citep{Tamai_Zingl_Georges_2019_PRX,Xiaoyu_Ru_PRL}. The experimental quasiparticle weights are $Z_{xz/yz}^{\text{exp}}\sim0.33$ and $Z_{xy}^{\text{exp}}\sim0.18$ for the three orbitals in the $t_{2g}$ subshell \citep{Tamai_Zingl_Georges_2019_PRX,Mackenzie_rev_2003}.

First, we construct the realistic tight-binding model from density functional theory (DFT)~\citep{Hohenberg1964,Kohn1965} and maximally localized Wannier function~\cite{Mazari2012}. We utilized Wien2k with 10000 k-points and the LDA functional for our DFT calculations
\citep{Wien2k}. {\color{black} The lattice parameters utilized in our DFT calculations are derived from the experimental structure at 100 K in Ref.~\onlinecite{SRO_struct}. The tetrahedron method is utilized for the Brillouin zone integration~\cite{tetrahedron}.} The wien2wannier and Wannier90 packages are then applied to construct the low-energy tight-binding model ~\citep{wien2wannier,wannier90}.
{\color{black} The low-energy tight-binding model is constructed from the Ru-$d_{xy}$, $d_{xz}$, and $d_{yz}$ orbitals with the energy window [$-3$ eV, $1$ eV] and $10\times10\times10$ k-points.} The spin-orbit coupling is ignored for simplicity. 
The Wannier tight-binding dispersion constructed from
DFT is shown in Fig. \ref{fig:SRO_DFT}. The $d_{xz}$ and $d_{yz}$ orbitals show a degenerate quasi-one dimensional density of states, and the $d_{xy}$ orbital has a two-dimensional dispersion with a van Hove singularity at the Fermi level.

\begin{figure}[t]
\begin{centering}
\includegraphics[scale=0.41]{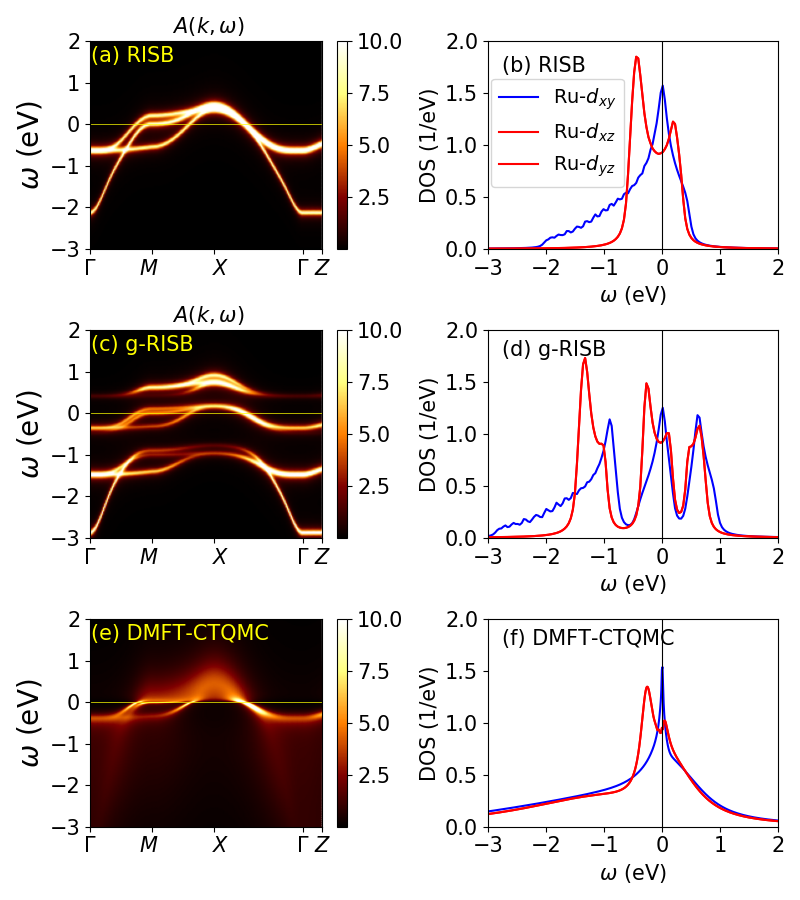}
\par\end{centering}
\caption{(a) The momentum-resolved spectral function $A(\mathbf{k},\omega)$ along the high-symmetry path in the Brillouin Zone and (b) the orbital-resolved density of states for Sr$_{2}$RuO$_{4}$ from RISB  with $U=2.3$ eV and $J=0.4$ eV. (c) The momentum-resolved spectral function $A(\mathbf{k},\omega)$ and (d) the orbital-resolved density of states from g-RISB with bath size $N_b=9$ at the same Coulomb parameters. (e) The momentum-resolved spectral function $A(\mathbf{k},\omega)$ and (f) the orbital-resolved density of states from DMFT with CTQMC solver with $U=2.3$ eV, $J=0.4$ eV, and $\beta=200$ eV$^{-1}$. {\color{black} A broadening factor $\eta=0.05$ is used in the density of states.} \label{fig:Akw_DOS_SRO}}
\end{figure}

\begin{figure}[h]
\begin{centering}
\includegraphics[scale=0.43]{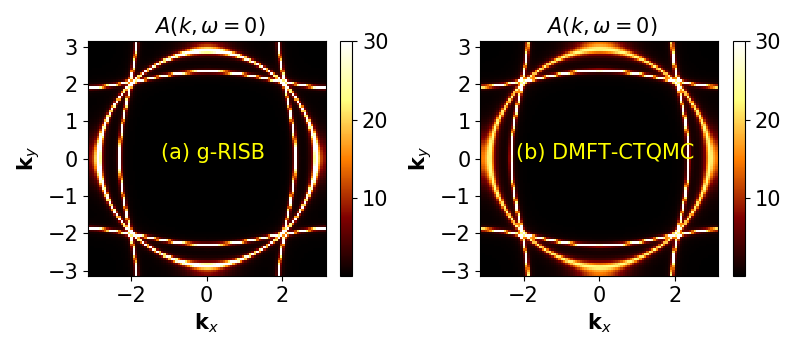}
\par\end{centering}
\caption{The momentum resolved spectral function for Sr$_2$RuO$_4$ at $\omega=0$, i.e., the Fermi surface, from (a) g-RISB with bath size $N_{b}=9$ for $U=2.3$ eV and $J=0.4$ eV and (b) DMFT with CTQMC solver for $U=2.3$ eV and $J=0.4$ eV and $\beta=200$ eV$^{-1}$. {\color{black} A broadening factor $\eta=0.05$ is used in the g-RISB density of states.}  \label{fig:SRO_FS}}
\end{figure}

Next, we apply RISB to include the electronic correlation effect. Figure \ref{fig:Akw_DOS_SRO} (a) and (b) show the momentum-resolved spectral function $A(\mathbf{k},\omega)$ along the high-symmetry path in the Brillouin Zone and the orbital-resolved density of states for RISB, respectively, at $U=2.3$ eV and $J=0.4$ eV. We found that the overall band structure is similar to DFT, with a slight renormalization of the bandwidth.
The quasiparticle weights for each orbital are $Z_{xz/yz}^{\text{RISB}}=0.67$ and $Z_{xy}^{\text{RISB}}=0.66$, which are inconsistent with the experiments. This behavior has been reported in Refs. \onlinecite{Facio_2018_PRB,Lechermann_2019_SRO}.

We now discuss the g-RISB spectral function and density of states shown in Fig. \ref{fig:Akw_DOS_SRO} (c) and (d), respectively. We use $N_{b}=9$ bath orbitals in our g-RISB calculations, so the total number of orbitals in the embedding Hamiltonian is $N_{\text{tot}}=12$, which can be efficiently solved by ED. The g-RISB band structure contains three groups of bands. The bands around the Fermi-level $\omega=0$ are the strongly renormalized quasiparticle bands. The quasiparticle weights are $Z_{xz/yz}^{\text{g-RISB}}=0.4$
and $Z_{xy}^{\text{g-RISB}}=0.35$ for each orbital, in reasonable agreement with experiments and the DMFT studies \citep{Mravlje_Ru_dmft,Minjae_2018_PRL,Tamai_Zingl_Georges_2019_PRX,Go_SRO_2020,Kugler2020,Cao_2021}.
The bands located at the energy windows {[}$-3$ eV, $-0.5$ eV{]} and {[}$0.2$ eV, $0.6$ eV{]} correspond to the lower and upper Hubbard bands, respectively. The DMFT results with CTQMC solver are shown in Fig. \ref{fig:Akw_DOS_SRO} (e) and (f), reproducing the previous studies {\color{black} with $Z^{\text{DMFT}}_{xz/yz}=0.3$ and $Z^{\text{DMFT}}_{xz/yz}=0.2$}~\citep{Kugler2020,Cao_2021}. Our results show that g-RISB
is able to accurately capture the low-energy quasiparticle bands around the Fermi level compared to DMFT. On the other hand, although the Hubbard bands in g-RISB locate at the correct energy scales, they do not have the incoherent feature in DMFT, with smeared dispersive bands. %This behavior is due to the fact that g-RISB is a mean-field approach that cannot capture the finite scattering rate in the imaginary part of the self-energy that generates the incoherent features in the spectral function. 
%{\color{black} This behavior is the consequence of the pole expansion form of the self-energy given in Eq.~\ref{eq:Sig}.}
We also show the Fermi surface in Fig.~\ref{fig:SRO_FS} for g-RISB with $N_b=9$ and DMFT with CTQMC solver. The g-RISB Fermi surface is in excellent agreement with DMFT.

\begin{figure}[t]
\begin{centering}
\includegraphics[scale=0.4]{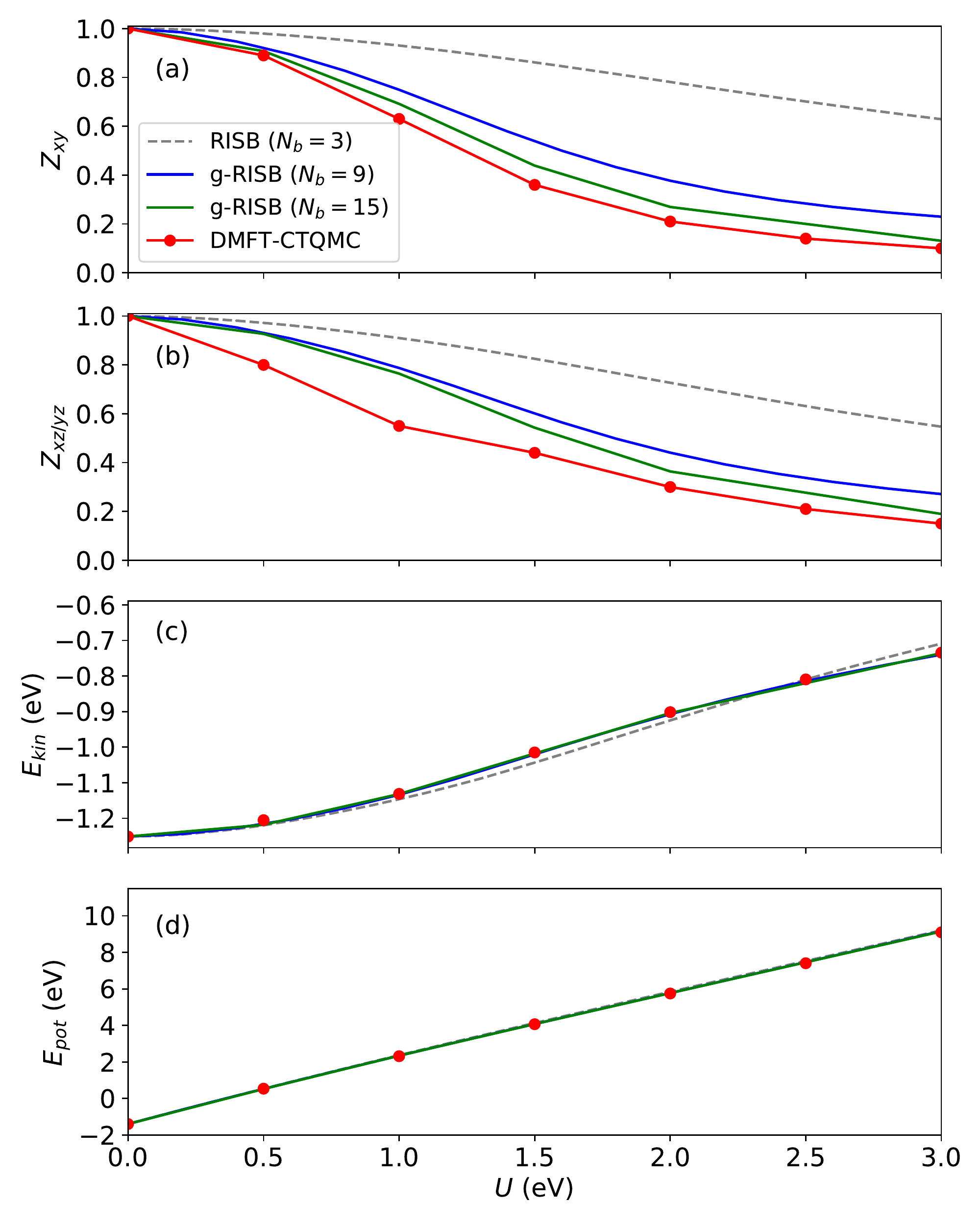}
\par\end{centering}
\caption{(a) The g-RISB and RISB quasiparticle weights for the $xy$ orbital $Z_{xy}$, (b) the $xz/yz$ orbital $Z_{xz/yz}$, and (c) the kinetic energy $E_{\text{kin}}$, and (d) the potential energy $E_{\text{pot}}$  for Sr$_2$RuO$_4$ as a function of Coulomb interaction $U$ with $J=0.2U$ and bath size $N_{b}=3,\ 9,\ 15$. The DMFT results with CTQMC solver at $\beta=200$ eV$^{-1}$ are shown for comparison. \label{fig:Z.pdf}}
\end{figure}

The quasiparticle weights for g-RISB, RISB, and DMFT as a function of Coulomb interaction $U$ with $J=0.2U$ are shown in Fig. \ref{fig:Z.pdf} (a) and (b). The RISB quasiparticle weights are overestimated even at large Coulomb interactions ($Z\sim0.6$ around $U=3.0$ eV) compared to the values observed in experiments ($Z_{xz/yz}^{\text{exp}}\sim0.33$ and $Z_{xy}^{\text{exp}}\sim0.18$). In addition, the quasiparticle weight in the $d_{xy}$-orbital is always larger than the $d_{xz/yz}$-orbitals for all the Coulomb interactions $U$, which is inconsistent with the experiments and the DMFT values.
On the other hand, g-RISB significantly improves the RISB quasiparticle weights toward the DMFT values for the considered interactions $U$, and the accuracy can be systematically improved with increasing $N_b$. The kinetic energy $E_{\text{kin}}$ and potential energy $E_{\text{pot}}$ are shown in Fig.~\ref{fig:Z.pdf} (c) and (d), respectively. We observed that RISB already provides accurate energy for this model, and the energy is almost indistinguishable from DMFT for g-RISB with $N_{b}=9$. To have a closer comparison with DMFT, we also provide the total energy values in Tab.~\ref{tab:energy} for the selected parameters.
%Around $U=2.5$ eV, the quasiparticle weights $Z_{xz/yz}^{\text{g-RISB}}=0.33$ and $Z_{xy}^{\text{g-RISB}}=0.28$ are in reasonable agreement with experimentally observed values.

%\hspace{2cm}
\begin{table}
\begin{tabular}{c c c c c c} 
 \hline
 U\;\;\;\; & $N_b=3$\;\; & $N_b=9$\;\; & $N_b=15$\;\;  & DMFT-CTQMC \\ [0.5ex] 
 \hline\hline
 1.5\;\; & 3.0895 & 3.056 & 3.055 & 3.053 \\ 
 \hline
 2.5\;\; & 6.718 & 6.649 & 6.646 & 6.648 \\
\hline
\end{tabular}
\caption{The g-RISB total energy $E_{\text{tot}}$ for the Sr$_2$RuO$_4$ model with different numbers of bath orbitals $N_b$ at $U=1.5$ eV and $2.5$  eV. The Hund's coupling is $J=0.2U$, and the electron filling is $n=4$. The DMFT energy at $\beta=200$ eV$^{-1}$ with the CTQMC solver is shown for comparison. The energy unit is in electron volts. \label{tab:energy}} 
\end{table}

%{\color{black}
\subsection{Discussions}

The numerical results presented here and in the previous studies on the one-orbital Hubbard model~\citep{gRISB_accuracy,gRISB_2017,Guerci_thesis,Guerci_TDgGA} 
strongly suggest that the total energy within g-RISB converges to the DMFT total energy from above as the number of ghost orbitals is increased. The g-RISB
spectral functions also approach the DMFT spectral functions as the number of bath orbitals in the embedding problem increases. For a given number of bath sites, we observe that g-RISB generally provides more accurate total energy, while DMFT-ED performs better for spectral properties.  An analogous situation occurs within electronic structure methods: density functional theory, which targets the exact density of the materials, provides a good approximation to the densities but is less precise for the spectral properties. On the other hand, the more expensive spectral density functional theory, which targets the spectra~\citep{DMFT_RMP_2006,Savrasov2004},  yields better results for spectral observables.
Likewise, we can view g-RISB as an approximation to an exact density matrix functional theory,  which targets the exact one body density matrix~\citep{Gilbert1975}, in the same spirit that DMFT can be viewed as an approximation to an exact spectral density functional~\citep{Savrasov2004,Chitra2000}.
%{\color{black} Another important observation regarding the convergence of g-RISB spectral properties is that the number of poles in the spectral function and self-energy in g-RISB is linearly proportional to the number of the ghost (bath) orbitals. Therefore, the convergence of the g-RISB spectral properties is than DMFT with the ED solver, where the number of the poles grows exponentially with the number of the bath orbitals.}
From these perspectives, we can understand why RISB gives good approximations to the total energy, while the spectra are clearly insufficient as the total spectral weight is too small,  and why adding more ghost sites in g-RISB rapidly improves the total energy. The expression of the g-RISB self-energy, which is similar to the  DMFT pole expansion (Eq.~\ref{eq:sig}), calls for further exploration of the connections between these two methods.
%}

%{\color{black} Add discussion here to discuss the fate of spectral function and imaginary part of the self-energy ... And also discuss that the scaling of the number of the self-energy is much slower than DMFT-ED. Therefore, g-RISB is by design more suitable for studying density matrix and less efficient for the spectral function..}

%\hspace{1.5cm}
\section{Conclusions}

We applied the g-RISB approach to the degenerate three-orbital Hubbard and a realistic Sr$_2$RuO$_4$ model and benchmarked its accuracy with DMFT. We provide numerical evidence that the accuracy of g-RISB is systematically improvable toward the exact DMFT limit in infinite dimensional multiorbital models with an increasing number of ghost orbitals. This allows a more precise description of Hund's metal behavior and Mott transition compared to the original RISB approach. 
%In addition, we observed the g-RISB energy is systematically more accurate than DMFT-ED with the same number of bath orbitals in the considered regimes of parameters. On the other hand, the g-RISB quasiparticle weight and the spectral functions require a larger number of bath orbitals to achieve similar accuracy to DMFT-ED. 
Moreover, we apply g-RISB to a realistic Sr$_2$RuO$_4$ model extracted from density functional theory and show that it produces reliable quasiparticle weights, Fermi surface, and {\color{black}low-energy} spectral function, compared to DMFT and experiments. 
{\color{black} Furthermore, we showcase the capability of employing the DMRG method as the impurity solver within the g-RISB framework, allowing us to explore systems with a greater number of ghost orbitals beyond the reach of the ED solver.}
%The DMRG method is utilized as an impurity solver within the g-RISB framework, allowing us to investigate systems with a larger number of ghost orbitals that the ED solver cannot access.}
%{\color{black} On the other hand, the incoherent high-energy Hubbard spectral properties are approximated by the coherent bands in g-RISB.} 
The connection between the g-RISB and the DMFT self-energy {\color{black} was} also discussed.
{\color{black}
%In addition, we discussed similarities and differences between the g-RISB and the DMFT, which we view as complementary approaches.  Unlike DMFT, which is a spectral density functional that targets spectral properties,  conceptually, g-RISB can be viewed as a  density matrix functional theory, targeting static correlation functions,  and indeed, these converge very well for a small number of bath sites.  Since g-RISB only requires the self-consistent computation of the single-particle density matrix, we demonstrated that this step can be carried out easily on a larger number of bath orbitals using DMRG.
%Our results demonstrate that g-RISB is a promising approach to strongly correlated materials that can be combined with first principle simulations.
%Notably,  gRISB can also be used to extract spectral information, but the convergence of their spectral properties will require more advanced tools.
}

Our results demonstrate that g-RISB is a promising approach to strongly correlated materials that can be combined with first principle simulations. 
%Moreover, since g-RISB only requires the self-consistent computation of the single-particle density matrix, which circumvents the calculation of Green's functions, it is expected to be more efficient than Green's function-based approaches, e.g., DMFT.
%In addition, the two-particle response functions and quasiparticle interaction vertices can be calculated from the recent development of RISB \citep{Lee_PRX_2021}. 
Future research will focus on the effect of spin-orbit coupling, the development of charge self-consistent DFT+g-RISB~\citep{DMFT_RMP_2006,Lanata_2015_PRX,Portobello}, the accuracy of the g-RISB response functions~\citep{Lee_PRX_2021}, and {\color{black} the potential of utilizing other tensor network approaches as a g-RISB impurity solver~\cite{Cao_2021,Bauernfeind2017}.}
%{\color{black} and the potential of g-RISB in describing the incoherent features with more bath orbitals where a more powerful tensor network solver has to be utilized~\cite{Cao_2021,Bauernfeind2017}}. %{\color{black} Our results also calls for further exploration on the convergence of g-RISB self-energy and spectral functions with the state of the arts DMRG and quantum Monte Carlo impurity solvers.}

\begin{acknowledgments}
We thank Antoine Georges for the useful discussions. T.-H. L, R.A., and G.K. were supported by the U.S. Department of Energy, Office of Science, Office of Advanced Scientific Computing Research, and Office of Basic Energy Sciences, Scientific Discovery through Advanced Computing (SciDAC) program under Award Number DE-SC0022198. This work was supported by the US Department of Energy, Office of Basic Energy Sciences as part of the Computation Material Science Program. N.L. gratefully acknowledges funding from the Simons Foundation (1030691, NL) and the Novo Nordisk Foundation through the Exploratory Interdisciplinary Synergy Programme project NNF19OC0057790.
\end{acknowledgments}

%\vspace{0.5cm}

\appendix

\begin{figure}[t]
\begin{centering}
\includegraphics[scale=0.4]{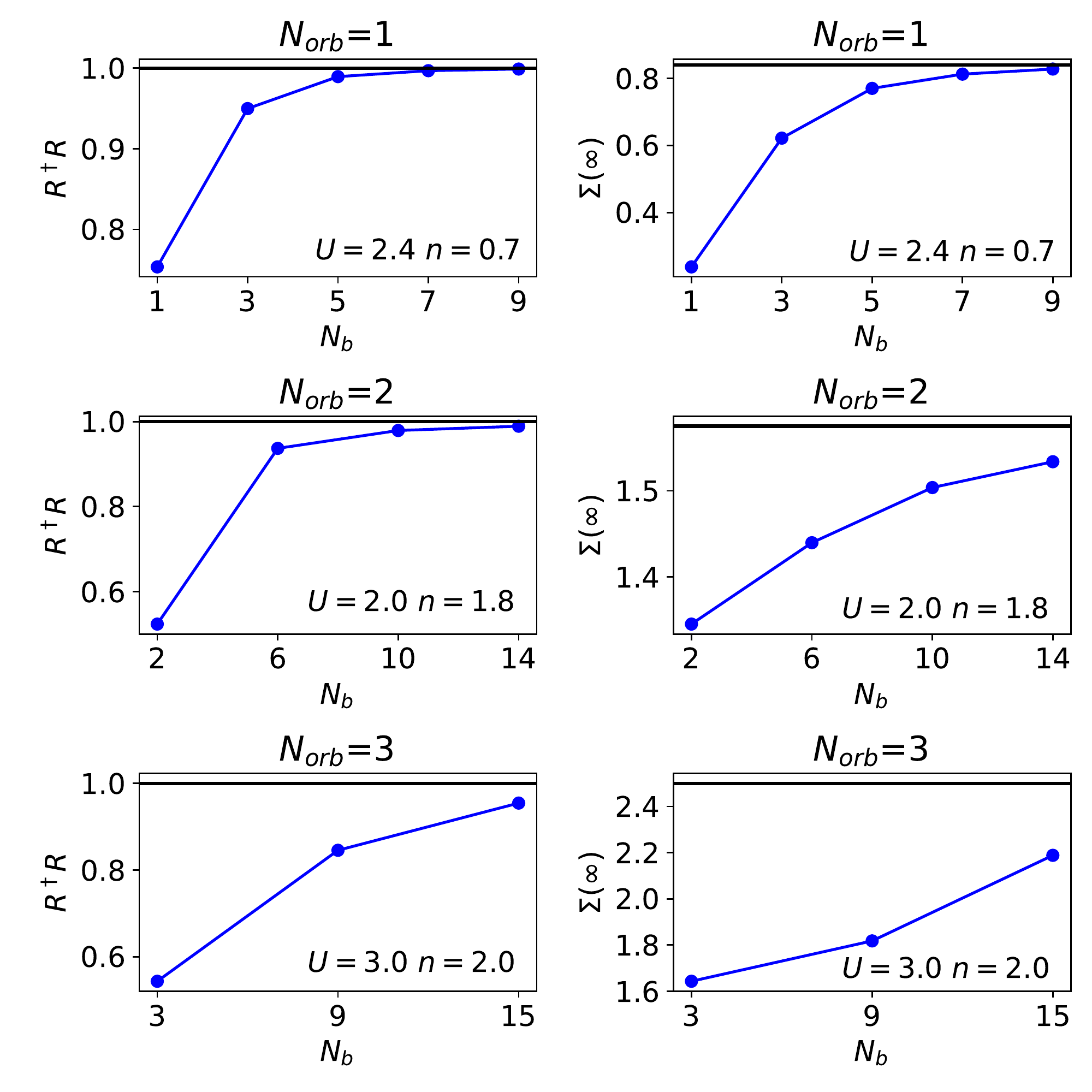}
\par\end{centering}
\caption{The diagonal element of $R^\dagger R$ and the self-energy at infinity frequency $\Sigma(\infty)$ as a function of the number of the bath orbitals $N_b$ for the degenerate one-orbital $N_{\text{orb}}=1$, two-orbital 
$N_{\text{orb}}=2$, and three-orbital $N_{\text{orb}}=3$ Hubbard model.  The Coulomb parameters $U$ and electron fillings $n$ are given in the figures. The $R^\dagger R$ and $\Sigma(\infty)$ approach one and the Hartree-Fock self-energy (the horizontal lines), respectively. \label{fig:RdR_Siginf}}
\end{figure}

{\color{black}
\section{Algorithm for solving g-RISB equations\label{sec:algorithm}}

Our algorithm for solving Eqs.~\ref{eq:SP1}-\ref{eq:SP6} is as follows: (1) starting from an initial guess of $R$ and $\lambda$, compute $\Delta$ from Eq.~\ref{eq:SP1}. (2) Solve $D$ from Eq.~\ref{eq:SP2}. (3) Solve $\lambda^c$ from Eq.~\ref{eq:SP3}. (4) Solve $|\Phi\rangle$ from Eq.~\ref{eq:SP4}. (5) Compute $\Delta$ from Eq.~\ref{eq:SP5}. (6) Compute the new $R$ from Eq.~\ref{eq:SP4} and the new $\lambda$ from Eq.~\ref{eq:SP3}. (7) Check the convergence of $R$ and $\lambda$. If not converged, go back to step (1). This algorithm is proposed in Ref.~\onlinecite{RISB_DMET_Ayral_2018}.
}

\section{Convergence of $R^\dagger R$ and $\Sigma(\infty)$\label{sec:RdR_Siginf}}

In this section, we discuss the behavior of the total spectral weight of the correlated degrees of freedom (computed from the variational parameters as $R^\dagger R$) and $\Sigma(\infty)$ as the number of the bath (ghost) orbitals $N_b$ increases. 

In Fig.~\ref{fig:RdR_Siginf}, we show the $R^\dagger R$ and $\Sigma(\infty)$ for the degenerate one-orbital $N_{\text{orb}}=1$, two-orbital $N_{\text{orb}}=2$, and three-orbital $N_{\text{orb}}=3$ Hubbard models for the selected Coulomb interactions and electron fillings. We use the Hubbard-Kanamori type of Coulomb interaction with Hund's coupling fixed at $J=0.25U$. Our results {\color{black} show} that $R^\dagger R$ approaches one and $\Sigma(\infty)$ approaches the Hartree-Fock self-energy, indicated by the horizontal lines, with increasing $N_b$. Note that $R^\dagger R$ and $\Sigma(\infty)$ are degenerate and diagonal matrices in the considered model.

%In this appendix, we discuss the convergence of $R^\dagger R$ and $\Sigma(\infty)$ as the number of the bath (ghost) orbitals $N_b$ increases. Note that $R^\dagger R=\tilde{R}^\dagger \tilde{R}$ due to the gauge symmetry (see Eq.~\ref{eq:gauge}). In Fig.~\ref{fig:RdR_Siginf}, we show the $R^\dagger R$ and $\Sigma(\infty)$ for the degenerate one-orbital $N_{\text{orb}}=1$, two-orbital $N_{\text{orb}}=2$, and three-orbital $N_{\text{orb}}=3$ Hubbard models for the selected Coulomb interactions and electron fillings. We use the Hubbard-Kanamori type of Coulomb interaction with Hund's coupling fixed at $J=0.25U$. Our results demonstrate that $R^\dagger R$ approaches one and $\Sigma(\infty)$ approaches the Hartree-Fock self-energy, indicated by the horizontal lines, with increasing $N_b$. Note that $R^\dagger R$ and $\Sigma(\infty)$ are degenerate and diagonal matrices in the considered model.

{\color{black}

%\begin{figure}[h]
%\begin{centering}
%\includegraphics[scale=0.35]{Sig_SRO}
%\par\end{centering}
%\caption{a-c) The real part of the g-RISB self-energy Re$\Sigma(\omega)$ for the one-orbital Hubbard model with increasing number of bath orbitals $N_b=3-15$. (d-f) The real part of the g-RISB self-energy Re$\Sigma(\omega)$ for the degenerate three-orbital Hubbard model with increasing number of bath orbitals $N_b=3-15$.  The DMFT-CTQMC self-energy (grey lines) with temperature $\beta=200$ is shown for comparison. A broadening factor $\eta=0.05$ is used and the energy unit is the half-bandwidth. \label{fig:ReSig_deg}}
%\end{figure}

\begin{figure}[t]
\begin{centering}
\includegraphics[scale=0.38]{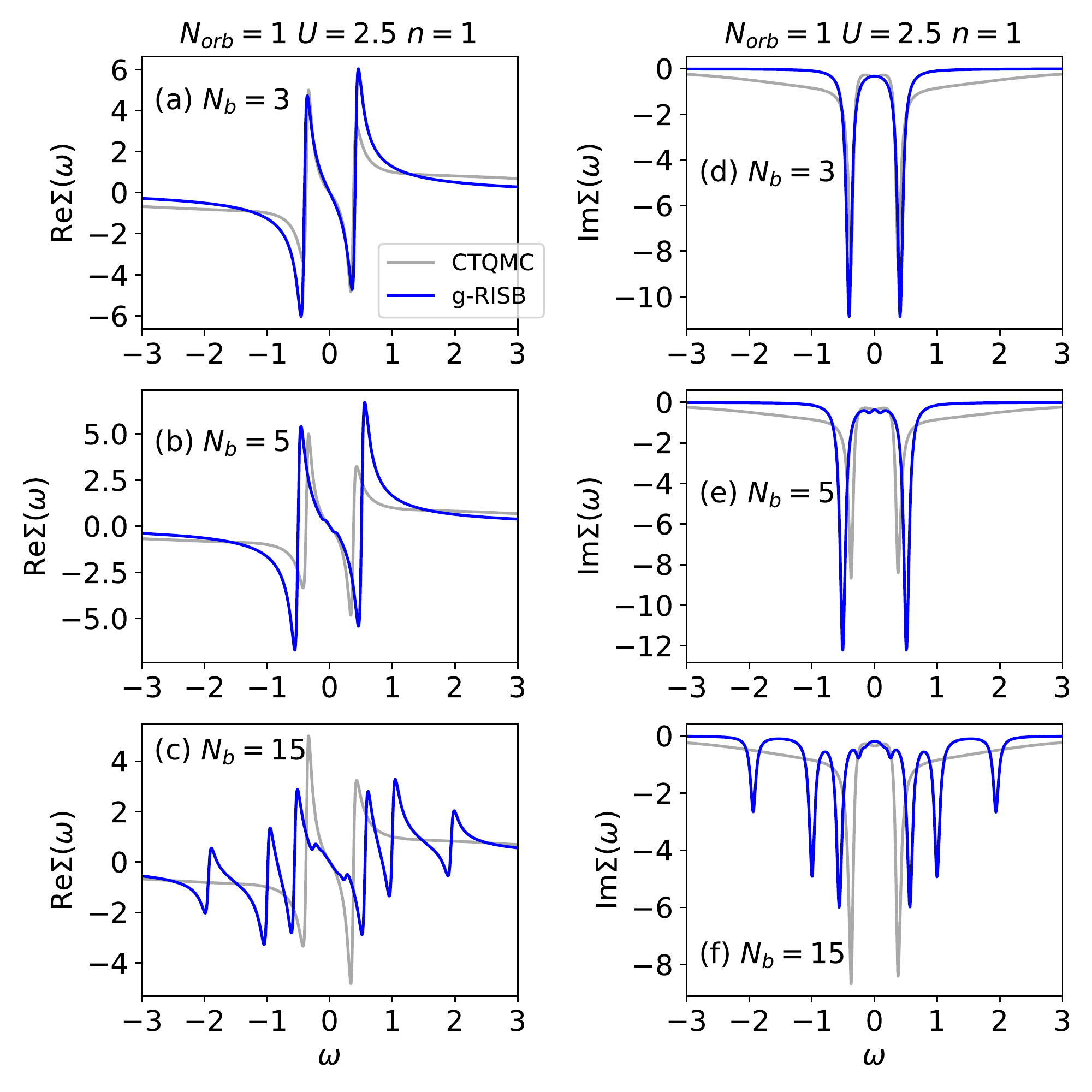}
\par\end{centering}
\caption{{\color{black} (a-c) The real part of the g-RISB self-energy Re$\Sigma(\omega)$ for the one-orbital Hubbard model with increasing number of bath orbitals $N_b=3-15$. (d-f) The imaginary part of the g-RISB self-energy Im$\Sigma(\omega)$ for the one-orbital Hubbard model with an increasing number of bath orbitals $N_b=3-15$. A broadening factor $\eta=0.05$ is used, and the energy unit is the half-bandwidth. The DMFT-CTQMC self-energy (grey lines) with temperature $\beta=200$ is shown for comparison.} \label{fig:Sig_deg_1orb}}
\end{figure}

\section{Self-energy comparison for the degenerate Hubbard models\label{sec:self-energy}}

%\subsection{Behavior of g-RISB self-energy on the real axis}\label{sec:self-energy}

\begin{figure}[t]
\begin{centering}
\includegraphics[scale=0.38]{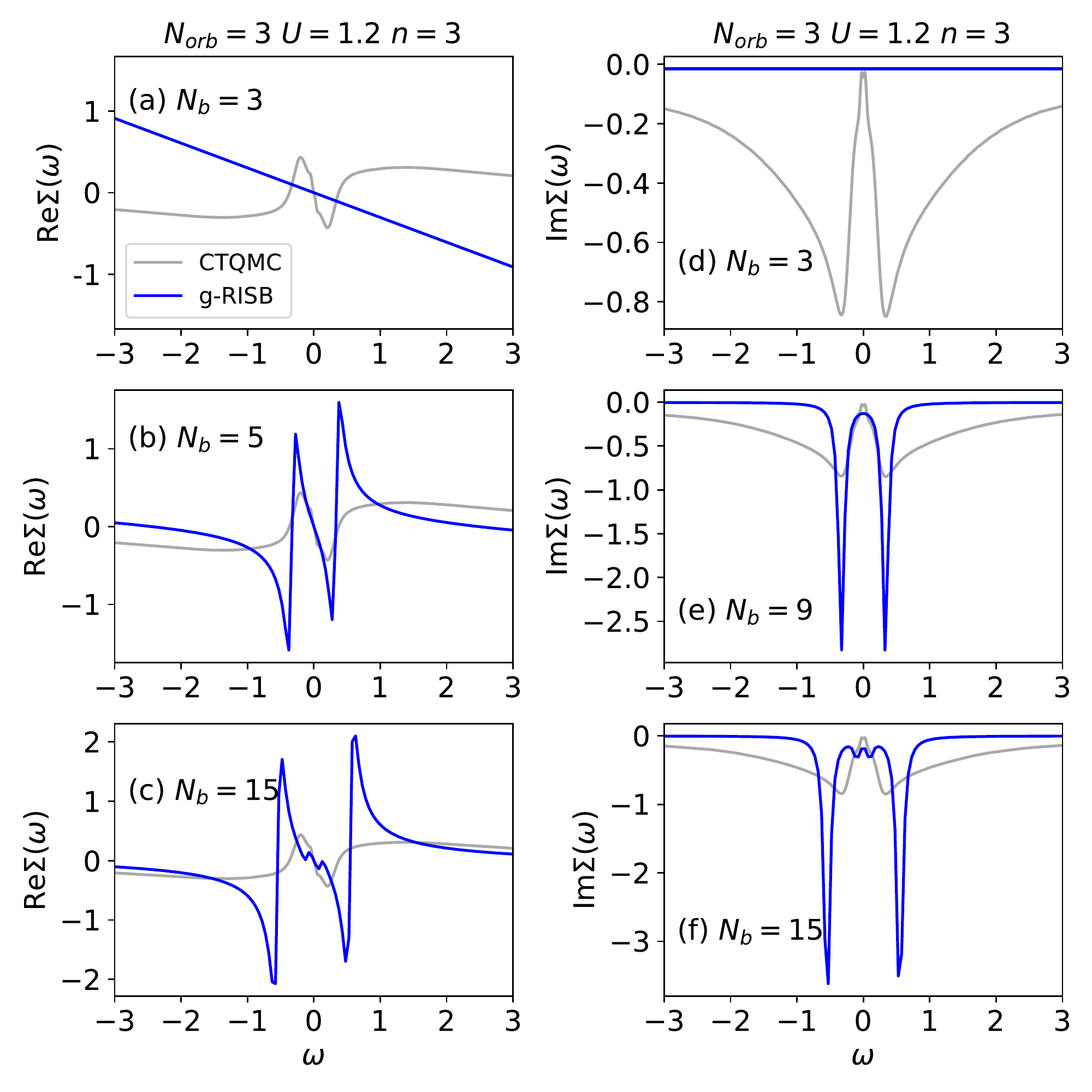}
\par\end{centering}
\caption{{\color{black}(a-c) The real part of the g-RISB self-energy Re$\Sigma(\omega)$ for the degenerate three-orbital Hubbard model with increasing number of bath orbitals $N_b=3-15$. (d-f) The imaginary part of the g-RISB self-energy Im$\Sigma(\omega)$ for the degenerate three-orbital Hubbard model with an increasing number of bath orbitals $N_b=3-15$. A broadening factor $\eta=0.05$ is used, and the energy unit is the half-bandwidth. The DMFT-CTQMC self-energy (grey lines) with temperature $\beta=200$ is shown for comparison.} \label{fig:Sig_deg_3orb}}
\end{figure}

The g-RISB self-energy for the one-orbital Hubbard model and the degenerate three-orbital Hubbard model are shown in Fig.~\ref{fig:Sig_deg_1orb} and Fig.~\ref{fig:Sig_deg_3orb}, respectively.  From our numerical results shown in Figs.~\ref{fig:Sig_deg_1orb} and \ref{fig:Sig_deg_3orb}, we observed that introducing more ghost (bath) orbitals in g-RISB leads to additional poles in the self-energy, as discussed in the main text. Although the current resolution with $N_b=15$ is not enough for us to conclude that the g-RISB self-energy converges to the exact DMFT limit, our results in one-orbital Hubbard model $N_{\text{orb}}=1$ indicates that the poles at higher energy are gradually captured with increasing $N_b$ and the asymptotic behavior of the Re$\Sigma(\omega)$ at high frequency is improved with increasing $N_b$.
These results suggest that the g-RISB self-energy may capture the incoherent feature with a large enough number of ghost orbitals and a proper broadening factor $\eta$.
For the degenerate three-orbital Hubbard model $N_{\text{orb}}=3$, the resolution with $N_b=15$ is insufficient for comparison with the DMFT-CTQMC self-energy and for drawing conclusions.
\begin{figure}[H]
\begin{centering}
\includegraphics[scale=0.375]{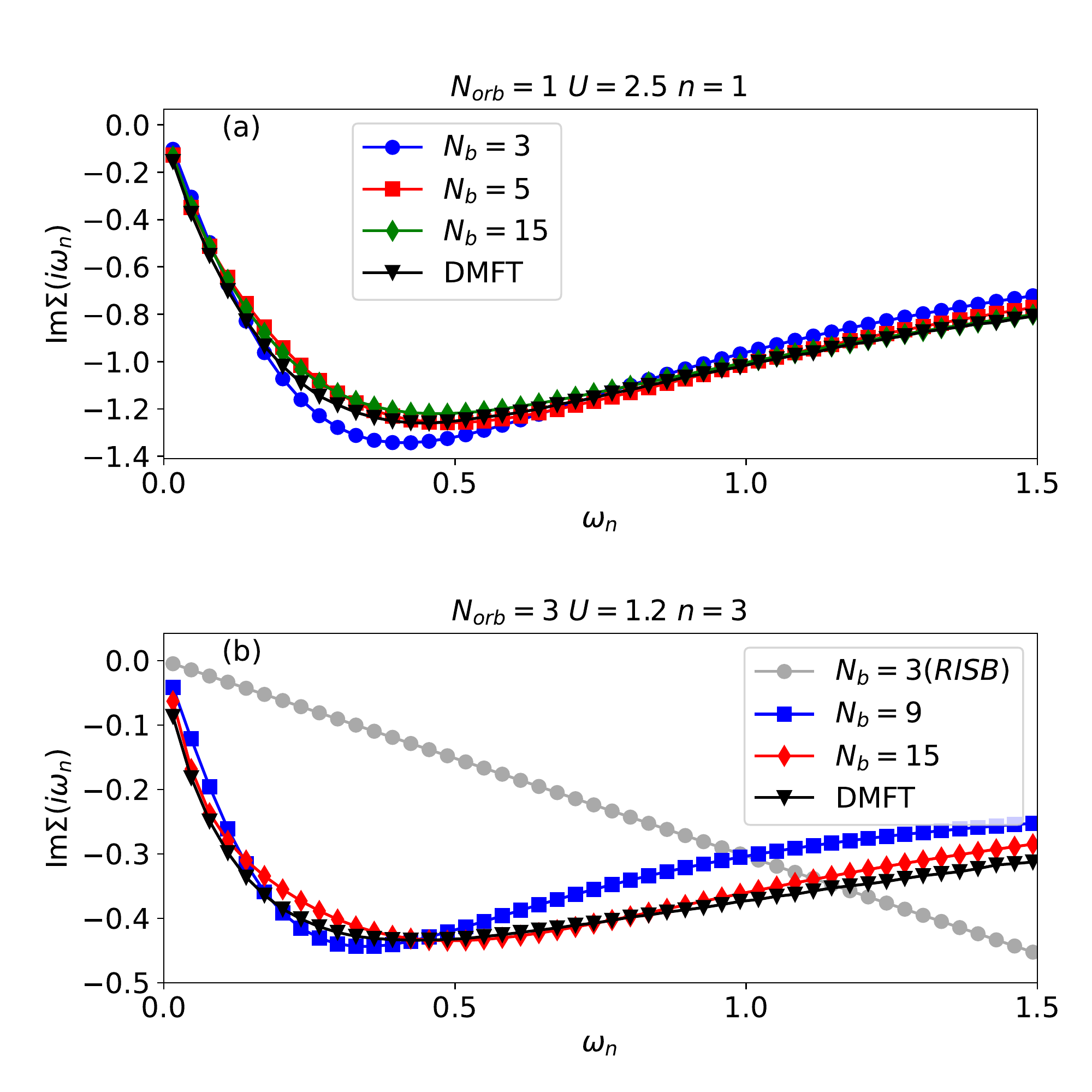}
\par\end{centering}
\caption{{\color{black} (a) The imaginary part of the g-RISB self-energy Im$\Sigma(i\omega_n)$ on the Matsubara axis for the one-orbital Hubbard model with increasing number of bath orbitals $N_b=3-15$. (b) The imaginary part of the g-RISB self-energy Im$\Sigma(\omega)$ for the degenerate three-orbital Hubbard model with an increasing number of bath orbitals $N_b=3-15$. The energy unit is the half-bandwidth. The DMFT-CTQMC self-energy (grey symbols) with temperature $\beta=200$ is shown for comparison.} \label{fig:ImSigiw_deg}}
\end{figure}
%We have also shown the self-energy on the Matsubara axis in Fig.~\ref{fig:ImSigiw_deg} at temperature $\beta=200$.

%\subsection{Behavior of the self-energy on the Matsubara axis}\label{sec:imaginary}

The g-RISB self-energy on the Matsubara axis, $z=i\omega_n $, is shown in Fig.~\ref{fig:ImSigiw_deg}(a) for the one-band Hubbard model with $U=2.5$ at half-filling $n=1$, and Fig.~\ref{fig:ImSigiw_deg}(b) for the degenerate three-orbital Hubbard model at $U=1.2$, $J=0.25U$, and half-filling $n=3$. We consider an infinite-dimensional Bethe lattice. With an increasing number of bath orbitals $N_b=3,\ 5,\ 15$, we observe systematic convergence of the self-energy toward the exact DMFT limit with an increasing number of bath orbitals in both models. Our calculations indicate that the self-energy at $N_b=15$ is in good quantitative agreement with the DMFT-CTQMC self-energy on the Matsubara axis, both at low and high frequencies. 

\section{Static observables at $J/U=0.1$ for the degenerate three-orbital Hubbard model \label{sec:JoverU0p1}}
To demonstrate the accuracy of g-RISB at different regimes of parameters, we show the static observables for the degenerate three-orbital Hubbard model with the ratio $J/U=0.1$ in Fig.~\ref{fig:benchmark_n2_JoverU0p1}. The overall accuracy is similar to the case with ratio $J/U=0.25$ shown in the main text Fig.~\ref{fig:benchmark_n2}.

\begin{figure}[H]
\begin{centering}
\includegraphics[scale=0.38]{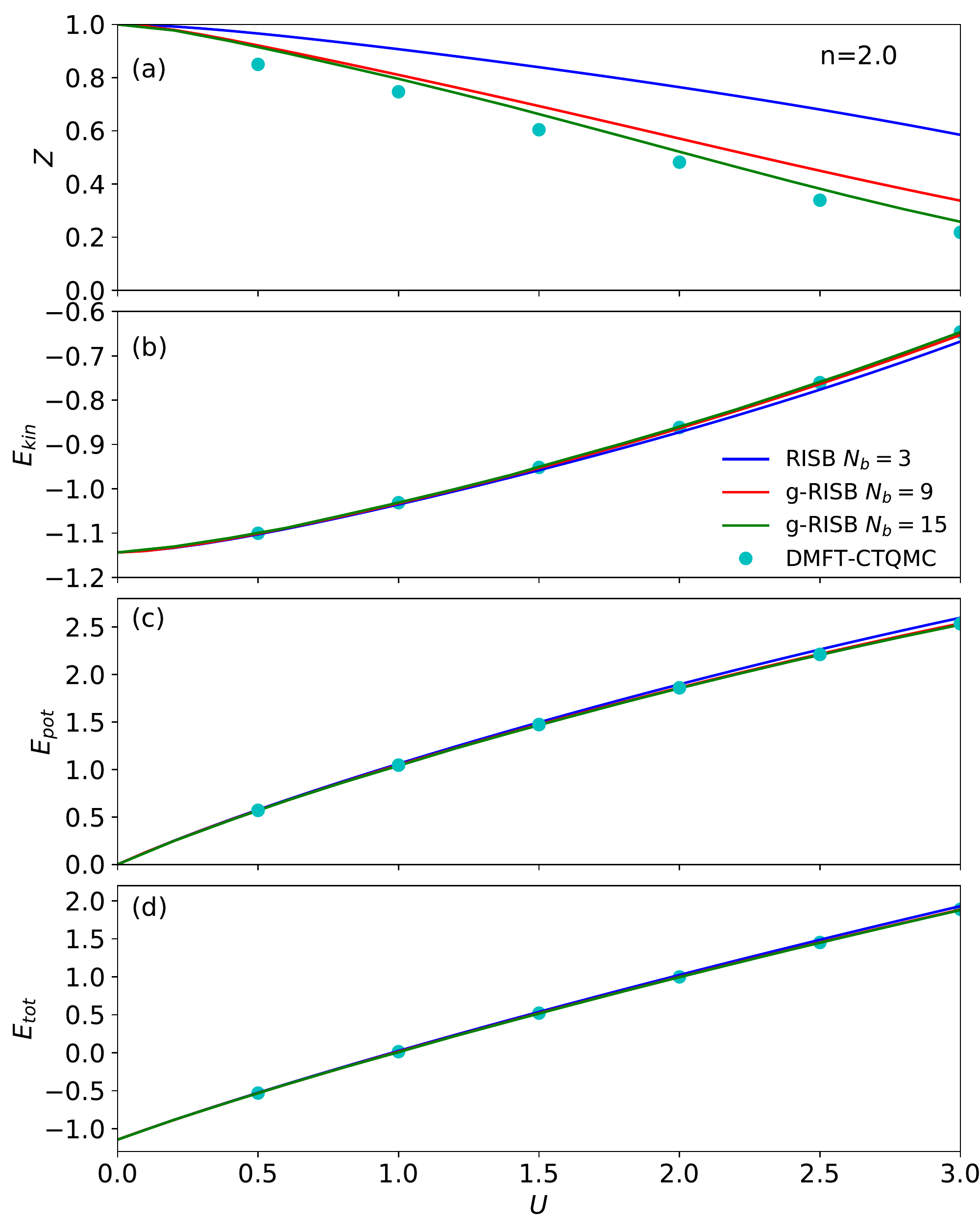}
\par\end{centering}
\caption{(a) The quasiparticle weights $Z$, (b) kinetic energy $E_{\text{kin}}$, (c) potential energy $E_{\text{pot}}$, and (d) total energy $E_{\text{tot}}$ for the degenerate three-orbital model on Bethe lattice for g-RISB with bath size $N_{b}=3,\ 9,\ 15$ and DMFT-ED with $N_{b}=3,\ 9$ as a function of Coulomb interaction $U$ with Hund's coupling $J=0.1U$ at filling $n=2$. The DMFT-CTQMC results at inverse temperature $\beta=200$ are shown for comparison. The energy unit is the half-bandwidth. \label{fig:benchmark_n2_JoverU0p1}}
\end{figure}

}

\clearpage

\bibliographystyle{apsrev}
\bibliography{ref}

\begin{thebibliography}{98}
\expandafter\ifx\csname natexlab\endcsname\relax\def\natexlab#1{#1}\fi
\expandafter\ifx\csname bibnamefont\endcsname\relax
  \def\bibnamefont#1{#1}\fi
\expandafter\ifx\csname bibfnamefont\endcsname\relax
  \def\bibfnamefont#1{#1}\fi
\expandafter\ifx\csname citenamefont\endcsname\relax
  \def\citenamefont#1{#1}\fi
\expandafter\ifx\csname url\endcsname\relax
  \def\url#1{\texttt{#1}}\fi
\expandafter\ifx\csname urlprefix\endcsname\relax\def\urlprefix{URL }\fi
\providecommand{\bibinfo}[2]{#2}
\providecommand{\eprint}[2][]{\url{#2}}

\bibitem[{\citenamefont{Lechermann et~al.}(2007)\citenamefont{Lechermann,
  Georges, Kotliar, and Parcollet}}]{Lecherman_2007}
\bibinfo{author}{\bibfnamefont{F.}~\bibnamefont{Lechermann}},
  \bibinfo{author}{\bibfnamefont{A.}~\bibnamefont{Georges}},
  \bibinfo{author}{\bibfnamefont{G.}~\bibnamefont{Kotliar}}, \bibnamefont{and}
  \bibinfo{author}{\bibfnamefont{O.}~\bibnamefont{Parcollet}},
  \bibinfo{journal}{Phys. Rev. B} \textbf{\bibinfo{volume}{76}},
  \bibinfo{pages}{155102} (\bibinfo{year}{2007}),
  \urlprefix\url{https://link.aps.org/doi/10.1103/PhysRevB.76.155102}.

\bibitem[{\citenamefont{Isidori and Capone}(2009)}]{Isidori_RISB_SC_2009}
\bibinfo{author}{\bibfnamefont{A.}~\bibnamefont{Isidori}} \bibnamefont{and}
  \bibinfo{author}{\bibfnamefont{M.}~\bibnamefont{Capone}},
  \bibinfo{journal}{Phys. Rev. B} \textbf{\bibinfo{volume}{80}},
  \bibinfo{pages}{115120} (\bibinfo{year}{2009}),
  \urlprefix\url{https://link.aps.org/doi/10.1103/PhysRevB.80.115120}.

\bibitem[{\citenamefont{Gutzwiller}(1963)}]{Gutzwiller1}
\bibinfo{author}{\bibfnamefont{M.~C.} \bibnamefont{Gutzwiller}},
  \bibinfo{journal}{Phys. Rev. Lett.} \textbf{\bibinfo{volume}{10}},
  \bibinfo{pages}{159} (\bibinfo{year}{1963}),
  \urlprefix\url{https://link.aps.org/doi/10.1103/PhysRevLett.10.159}.

\bibitem[{\citenamefont{Gutzwiller}(1964)}]{Gutzwiller2}
\bibinfo{author}{\bibfnamefont{M.~C.} \bibnamefont{Gutzwiller}},
  \bibinfo{journal}{Phys. Rev.} \textbf{\bibinfo{volume}{134}},
  \bibinfo{pages}{A923} (\bibinfo{year}{1964}),
  \urlprefix\url{https://link.aps.org/doi/10.1103/PhysRev.134.A923}.

\bibitem[{\citenamefont{Metzner and Vollhardt}(1989)}]{Metzner_Vollhardt_1989}
\bibinfo{author}{\bibfnamefont{W.}~\bibnamefont{Metzner}} \bibnamefont{and}
  \bibinfo{author}{\bibfnamefont{D.}~\bibnamefont{Vollhardt}},
  \bibinfo{journal}{Phys. Rev. Lett.} \textbf{\bibinfo{volume}{62}},
  \bibinfo{pages}{324} (\bibinfo{year}{1989}),
  \urlprefix\url{https://link.aps.org/doi/10.1103/PhysRevLett.62.324}.

\bibitem[{\citenamefont{B\"unemann et~al.}(1998)\citenamefont{B\"unemann,
  Weber, and Gebhard}}]{Bunemann_mulorb_GA}
\bibinfo{author}{\bibfnamefont{J.}~\bibnamefont{B\"unemann}},
  \bibinfo{author}{\bibfnamefont{W.}~\bibnamefont{Weber}}, \bibnamefont{and}
  \bibinfo{author}{\bibfnamefont{F.}~\bibnamefont{Gebhard}},
  \bibinfo{journal}{Phys. Rev. B} \textbf{\bibinfo{volume}{57}},
  \bibinfo{pages}{6896} (\bibinfo{year}{1998}),
  \urlprefix\url{https://link.aps.org/doi/10.1103/PhysRevB.57.6896}.

\bibitem[{\citenamefont{B\"unemann and Gebhard}(2007)}]{Bunemann_2007}
\bibinfo{author}{\bibfnamefont{J.}~\bibnamefont{B\"unemann}} \bibnamefont{and}
  \bibinfo{author}{\bibfnamefont{F.}~\bibnamefont{Gebhard}},
  \bibinfo{journal}{Phys. Rev. B} \textbf{\bibinfo{volume}{76}},
  \bibinfo{pages}{193104} (\bibinfo{year}{2007}),
  \urlprefix\url{https://link.aps.org/doi/10.1103/PhysRevB.76.193104}.

\bibitem[{\citenamefont{Fabrizio}(2007)}]{Fabrizio_SC}
\bibinfo{author}{\bibfnamefont{M.}~\bibnamefont{Fabrizio}},
  \bibinfo{journal}{Phys. Rev. B} \textbf{\bibinfo{volume}{76}},
  \bibinfo{pages}{165110} (\bibinfo{year}{2007}),
  \urlprefix\url{https://link.aps.org/doi/10.1103/PhysRevB.76.165110}.

\bibitem[{\citenamefont{Lanat\`a et~al.}(2008)\citenamefont{Lanat\`a, Barone,
  and Fabrizio}}]{Lanata_2008}
\bibinfo{author}{\bibfnamefont{N.}~\bibnamefont{Lanat\`a}},
  \bibinfo{author}{\bibfnamefont{P.}~\bibnamefont{Barone}}, \bibnamefont{and}
  \bibinfo{author}{\bibfnamefont{M.}~\bibnamefont{Fabrizio}},
  \bibinfo{journal}{Phys. Rev. B} \textbf{\bibinfo{volume}{78}},
  \bibinfo{pages}{155127} (\bibinfo{year}{2008}),
  \urlprefix\url{https://link.aps.org/doi/10.1103/PhysRevB.78.155127}.

\bibitem[{\citenamefont{Kotliar and Ruckenstein}(1986)}]{Kotliar1986}
\bibinfo{author}{\bibfnamefont{G.}~\bibnamefont{Kotliar}} \bibnamefont{and}
  \bibinfo{author}{\bibfnamefont{A.~E.} \bibnamefont{Ruckenstein}},
  \bibinfo{journal}{Phys. Rev. Lett.} \textbf{\bibinfo{volume}{57}},
  \bibinfo{pages}{1362} (\bibinfo{year}{1986}),
  \urlprefix\url{https://link.aps.org/doi/10.1103/PhysRevLett.57.1362}.

\bibitem[{\citenamefont{Li et~al.}(1991)\citenamefont{Li, Sun, and
  W{\"o}lfle}}]{Li1991}
\bibinfo{author}{\bibfnamefont{T.}~\bibnamefont{Li}},
  \bibinfo{author}{\bibfnamefont{Y.~S.} \bibnamefont{Sun}}, \bibnamefont{and}
  \bibinfo{author}{\bibfnamefont{P.}~\bibnamefont{W{\"o}lfle}},
  \bibinfo{journal}{Zeitschrift f{\"u}r Physik B Condensed Matter}
  \textbf{\bibinfo{volume}{82}}, \bibinfo{pages}{369} (\bibinfo{year}{1991}),
  ISSN \bibinfo{issn}{1431-584X},
  \urlprefix\url{https://doi.org/10.1007/BF01357181}.

\bibitem[{\citenamefont{de'Medici et~al.}(2005)\citenamefont{de'Medici,
  Georges, and Biermann}}]{Medici2005}
\bibinfo{author}{\bibfnamefont{L.}~\bibnamefont{de'Medici}},
  \bibinfo{author}{\bibfnamefont{A.}~\bibnamefont{Georges}}, \bibnamefont{and}
  \bibinfo{author}{\bibfnamefont{S.}~\bibnamefont{Biermann}},
  \bibinfo{journal}{Phys. Rev. B} \textbf{\bibinfo{volume}{72}},
  \bibinfo{pages}{205124} (\bibinfo{year}{2005}),
  \urlprefix\url{https://link.aps.org/doi/10.1103/PhysRevB.72.205124}.

\bibitem[{\citenamefont{Crispino et~al.}(2023)\citenamefont{Crispino,
  Chatzieleftheriou, Gorni, and de' Medici}}]{Matteo_SS_2023}
\bibinfo{author}{\bibfnamefont{M.}~\bibnamefont{Crispino}},
  \bibinfo{author}{\bibfnamefont{M.}~\bibnamefont{Chatzieleftheriou}},
  \bibinfo{author}{\bibfnamefont{T.}~\bibnamefont{Gorni}}, \bibnamefont{and}
  \bibinfo{author}{\bibfnamefont{L.}~\bibnamefont{de' Medici}},
  \bibinfo{journal}{Phys. Rev. B} \textbf{\bibinfo{volume}{107}},
  \bibinfo{pages}{155149} (\bibinfo{year}{2023}),
  \urlprefix\url{https://link.aps.org/doi/10.1103/PhysRevB.107.155149}.

\bibitem[{\citenamefont{Yu and Si}(2012)}]{Yu2012}
\bibinfo{author}{\bibfnamefont{R.}~\bibnamefont{Yu}} \bibnamefont{and}
  \bibinfo{author}{\bibfnamefont{Q.}~\bibnamefont{Si}}, \bibinfo{journal}{Phys.
  Rev. B} \textbf{\bibinfo{volume}{86}}, \bibinfo{pages}{085104}
  (\bibinfo{year}{2012}),
  \urlprefix\url{https://link.aps.org/doi/10.1103/PhysRevB.86.085104}.

\bibitem[{\citenamefont{Florens and Georges}(2002)}]{Serge_rotor_2002}
\bibinfo{author}{\bibfnamefont{S.}~\bibnamefont{Florens}} \bibnamefont{and}
  \bibinfo{author}{\bibfnamefont{A.}~\bibnamefont{Georges}},
  \bibinfo{journal}{Phys. Rev. B} \textbf{\bibinfo{volume}{66}},
  \bibinfo{pages}{165111} (\bibinfo{year}{2002}),
  \urlprefix\url{https://link.aps.org/doi/10.1103/PhysRevB.66.165111}.

\bibitem[{\citenamefont{Florens and Georges}(2004)}]{Serge_rotor_2004}
\bibinfo{author}{\bibfnamefont{S.}~\bibnamefont{Florens}} \bibnamefont{and}
  \bibinfo{author}{\bibfnamefont{A.}~\bibnamefont{Georges}},
  \bibinfo{journal}{Phys. Rev. B} \textbf{\bibinfo{volume}{70}},
  \bibinfo{pages}{035114} (\bibinfo{year}{2004}),
  \urlprefix\url{https://link.aps.org/doi/10.1103/PhysRevB.70.035114}.

\bibitem[{\citenamefont{Georgescu and Ismail-Beigi}(2015)}]{Georgescu2015}
\bibinfo{author}{\bibfnamefont{A.~B.} \bibnamefont{Georgescu}}
  \bibnamefont{and}
  \bibinfo{author}{\bibfnamefont{S.}~\bibnamefont{Ismail-Beigi}},
  \bibinfo{journal}{Phys. Rev. B} \textbf{\bibinfo{volume}{92}},
  \bibinfo{pages}{235117} (\bibinfo{year}{2015}),
  \urlprefix\url{https://link.aps.org/doi/10.1103/PhysRevB.92.235117}.

\bibitem[{\citenamefont{Koga et~al.}(2004)\citenamefont{Koga, Kawakami, Rice,
  and Sigrist}}]{Koga_OSMT}
\bibinfo{author}{\bibfnamefont{A.}~\bibnamefont{Koga}},
  \bibinfo{author}{\bibfnamefont{N.}~\bibnamefont{Kawakami}},
  \bibinfo{author}{\bibfnamefont{T.~M.} \bibnamefont{Rice}}, \bibnamefont{and}
  \bibinfo{author}{\bibfnamefont{M.}~\bibnamefont{Sigrist}},
  \bibinfo{journal}{Phys. Rev. Lett.} \textbf{\bibinfo{volume}{92}},
  \bibinfo{pages}{216402} (\bibinfo{year}{2004}),
  \urlprefix\url{https://link.aps.org/doi/10.1103/PhysRevLett.92.216402}.

\bibitem[{\citenamefont{Georges et~al.}(2013)\citenamefont{Georges, de' Medici,
  and Mravlje}}]{Georges_Hunds_review}
\bibinfo{author}{\bibfnamefont{A.}~\bibnamefont{Georges}},
  \bibinfo{author}{\bibfnamefont{L.}~\bibnamefont{de' Medici}},
  \bibnamefont{and} \bibinfo{author}{\bibfnamefont{J.}~\bibnamefont{Mravlje}},
  \bibinfo{journal}{Annual Review of Condensed Matter Physics}
  \textbf{\bibinfo{volume}{4}}, \bibinfo{pages}{137} (\bibinfo{year}{2013}),
  \urlprefix\url{https://doi.org/10.1146/annurev-conmatphys-020911-125045}.

\bibitem[{\citenamefont{de' Medici et~al.}(2011)\citenamefont{de' Medici,
  Mravlje, and Georges}}]{Medici_Janus_PRL}
\bibinfo{author}{\bibfnamefont{L.}~\bibnamefont{de' Medici}},
  \bibinfo{author}{\bibfnamefont{J.}~\bibnamefont{Mravlje}}, \bibnamefont{and}
  \bibinfo{author}{\bibfnamefont{A.}~\bibnamefont{Georges}},
  \bibinfo{journal}{Phys. Rev. Lett.} \textbf{\bibinfo{volume}{107}},
  \bibinfo{pages}{256401} (\bibinfo{year}{2011}),
  \urlprefix\url{https://link.aps.org/doi/10.1103/PhysRevLett.107.256401}.

\bibitem[{\citenamefont{Haule and Kotliar}(2009)}]{HauleHunds}
\bibinfo{author}{\bibfnamefont{K.}~\bibnamefont{Haule}} \bibnamefont{and}
  \bibinfo{author}{\bibfnamefont{G.}~\bibnamefont{Kotliar}},
  \bibinfo{journal}{New Journal of Physics} \textbf{\bibinfo{volume}{11}},
  \bibinfo{pages}{025021} (\bibinfo{year}{2009}),
  \urlprefix\url{http://stacks.iop.org/1367-2630/11/i=2/a=025021}.

\bibitem[{\citenamefont{Fanfarillo and Bascones}(2015)}]{Fanfarillo_2015}
\bibinfo{author}{\bibfnamefont{L.}~\bibnamefont{Fanfarillo}} \bibnamefont{and}
  \bibinfo{author}{\bibfnamefont{E.}~\bibnamefont{Bascones}},
  \bibinfo{journal}{Phys. Rev. B} \textbf{\bibinfo{volume}{92}},
  \bibinfo{pages}{075136} (\bibinfo{year}{2015}),
  \urlprefix\url{https://link.aps.org/doi/10.1103/PhysRevB.92.075136}.

\bibitem[{\citenamefont{Isidori et~al.}(2019)\citenamefont{Isidori,
  Berovi\ifmmode~\acute{c}\else \'{c}\fi{}, Fanfarillo, de' Medici, Fabrizio,
  and Capone}}]{Isidori_2019_PRL}
\bibinfo{author}{\bibfnamefont{A.}~\bibnamefont{Isidori}},
  \bibinfo{author}{\bibfnamefont{M.}~\bibnamefont{Berovi\ifmmode~\acute{c}\else
  \'{c}\fi{}}}, \bibinfo{author}{\bibfnamefont{L.}~\bibnamefont{Fanfarillo}},
  \bibinfo{author}{\bibfnamefont{L.}~\bibnamefont{de' Medici}},
  \bibinfo{author}{\bibfnamefont{M.}~\bibnamefont{Fabrizio}}, \bibnamefont{and}
  \bibinfo{author}{\bibfnamefont{M.}~\bibnamefont{Capone}},
  \bibinfo{journal}{Phys. Rev. Lett.} \textbf{\bibinfo{volume}{122}},
  \bibinfo{pages}{186401} (\bibinfo{year}{2019}),
  \urlprefix\url{https://link.aps.org/doi/10.1103/PhysRevLett.122.186401}.

\bibitem[{\citenamefont{Georges et~al.}(1996)\citenamefont{Georges, Kotliar,
  Krauth, and Rozenberg}}]{DMFT_RMP_1996}
\bibinfo{author}{\bibfnamefont{A.}~\bibnamefont{Georges}},
  \bibinfo{author}{\bibfnamefont{G.}~\bibnamefont{Kotliar}},
  \bibinfo{author}{\bibfnamefont{W.}~\bibnamefont{Krauth}}, \bibnamefont{and}
  \bibinfo{author}{\bibfnamefont{M.~J.} \bibnamefont{Rozenberg}},
  \bibinfo{journal}{Rev. Mod. Phys.} \textbf{\bibinfo{volume}{68}},
  \bibinfo{pages}{13} (\bibinfo{year}{1996}),
  \urlprefix\url{https://link.aps.org/doi/10.1103/RevModPhys.68.13}.

\bibitem[{\citenamefont{Deng et~al.}(2009)\citenamefont{Deng, Wang, Dai, and
  Fang}}]{Deng_2009}
\bibinfo{author}{\bibfnamefont{X.}~\bibnamefont{Deng}},
  \bibinfo{author}{\bibfnamefont{L.}~\bibnamefont{Wang}},
  \bibinfo{author}{\bibfnamefont{X.}~\bibnamefont{Dai}}, \bibnamefont{and}
  \bibinfo{author}{\bibfnamefont{Z.}~\bibnamefont{Fang}},
  \bibinfo{journal}{Phys. Rev. B} \textbf{\bibinfo{volume}{79}},
  \bibinfo{pages}{075114} (\bibinfo{year}{2009}),
  \urlprefix\url{https://link.aps.org/doi/10.1103/PhysRevB.79.075114}.

\bibitem[{\citenamefont{Lanat\`a et~al.}(2012)\citenamefont{Lanat\`a, Strand,
  Dai, and Hellsing}}]{Lanata_2012}
\bibinfo{author}{\bibfnamefont{N.}~\bibnamefont{Lanat\`a}},
  \bibinfo{author}{\bibfnamefont{H.~U.~R.} \bibnamefont{Strand}},
  \bibinfo{author}{\bibfnamefont{X.}~\bibnamefont{Dai}}, \bibnamefont{and}
  \bibinfo{author}{\bibfnamefont{B.}~\bibnamefont{Hellsing}},
  \bibinfo{journal}{Phys. Rev. B} \textbf{\bibinfo{volume}{85}},
  \bibinfo{pages}{035133} (\bibinfo{year}{2012}),
  \urlprefix\url{https://link.aps.org/doi/10.1103/PhysRevB.85.035133}.

\bibitem[{\citenamefont{Lanat\`a et~al.}(2013)\citenamefont{Lanat\`a, Yao,
  Wang, Ho, Schmalian, Haule, and Kotliar}}]{Lanata_2013_Ce}
\bibinfo{author}{\bibfnamefont{N.}~\bibnamefont{Lanat\`a}},
  \bibinfo{author}{\bibfnamefont{Y.-X.} \bibnamefont{Yao}},
  \bibinfo{author}{\bibfnamefont{C.-Z.} \bibnamefont{Wang}},
  \bibinfo{author}{\bibfnamefont{K.-M.} \bibnamefont{Ho}},
  \bibinfo{author}{\bibfnamefont{J.}~\bibnamefont{Schmalian}},
  \bibinfo{author}{\bibfnamefont{K.}~\bibnamefont{Haule}}, \bibnamefont{and}
  \bibinfo{author}{\bibfnamefont{G.}~\bibnamefont{Kotliar}},
  \bibinfo{journal}{Phys. Rev. Lett.} \textbf{\bibinfo{volume}{111}},
  \bibinfo{pages}{196801} (\bibinfo{year}{2013}),
  \urlprefix\url{https://link.aps.org/doi/10.1103/PhysRevLett.111.196801}.

\bibitem[{\citenamefont{Lanat\`a et~al.}(2015)\citenamefont{Lanat\`a, Yao,
  Wang, Ho, and Kotliar}}]{Lanata_2015_PRX}
\bibinfo{author}{\bibfnamefont{N.}~\bibnamefont{Lanat\`a}},
  \bibinfo{author}{\bibfnamefont{Y.}~\bibnamefont{Yao}},
  \bibinfo{author}{\bibfnamefont{C.-Z.} \bibnamefont{Wang}},
  \bibinfo{author}{\bibfnamefont{K.-M.} \bibnamefont{Ho}}, \bibnamefont{and}
  \bibinfo{author}{\bibfnamefont{G.}~\bibnamefont{Kotliar}},
  \bibinfo{journal}{Phys. Rev. X} \textbf{\bibinfo{volume}{5}},
  \bibinfo{pages}{011008} (\bibinfo{year}{2015}),
  \urlprefix\url{https://link.aps.org/doi/10.1103/PhysRevX.5.011008}.

\bibitem[{\citenamefont{Lanat\`a
  et~al.}(2017{\natexlab{a}})\citenamefont{Lanat\`a, Yao, Deng,
  Dobrosavljevi\ifmmode~\acute{c}\else \'{c}\fi{}, and
  Kotliar}}]{Lanata_2017_PRL}
\bibinfo{author}{\bibfnamefont{N.}~\bibnamefont{Lanat\`a}},
  \bibinfo{author}{\bibfnamefont{Y.}~\bibnamefont{Yao}},
  \bibinfo{author}{\bibfnamefont{X.}~\bibnamefont{Deng}},
  \bibinfo{author}{\bibfnamefont{V.}~\bibnamefont{Dobrosavljevi\ifmmode~\acute{c}\else
  \'{c}\fi{}}}, \bibnamefont{and}
  \bibinfo{author}{\bibfnamefont{G.}~\bibnamefont{Kotliar}},
  \bibinfo{journal}{Phys. Rev. Lett.} \textbf{\bibinfo{volume}{118}},
  \bibinfo{pages}{126401} (\bibinfo{year}{2017}{\natexlab{a}}),
  \urlprefix\url{https://link.aps.org/doi/10.1103/PhysRevLett.118.126401}.

\bibitem[{\citenamefont{Lanat{\`a} et~al.}(2019)\citenamefont{Lanat{\`a}, Lee,
  Yao, Stevanovi{\'{c}}, and Dobrosavljevi{\'{c}}}}]{Lanata2019_NPJ}
\bibinfo{author}{\bibfnamefont{N.}~\bibnamefont{Lanat{\`a}}},
  \bibinfo{author}{\bibfnamefont{T.-H.} \bibnamefont{Lee}},
  \bibinfo{author}{\bibfnamefont{Y.-X.} \bibnamefont{Yao}},
  \bibinfo{author}{\bibfnamefont{V.}~\bibnamefont{Stevanovi{\'{c}}}},
  \bibnamefont{and}
  \bibinfo{author}{\bibfnamefont{V.}~\bibnamefont{Dobrosavljevi{\'{c}}}},
  \bibinfo{journal}{npj Computational Materials} \textbf{\bibinfo{volume}{5}},
  \bibinfo{pages}{30} (\bibinfo{year}{2019}), ISSN \bibinfo{issn}{2057-3960},
  \urlprefix\url{https://doi.org/10.1038/s41524-019-0169-0}.

\bibitem[{\citenamefont{Piefke and Lechermann}(2011)}]{Christoph_2011}
\bibinfo{author}{\bibfnamefont{C.}~\bibnamefont{Piefke}} \bibnamefont{and}
  \bibinfo{author}{\bibfnamefont{F.}~\bibnamefont{Lechermann}},
  \bibinfo{journal}{physica status solidi (b)} \textbf{\bibinfo{volume}{248}},
  \bibinfo{pages}{2269} (\bibinfo{year}{2011}),
  \eprint{https://onlinelibrary.wiley.com/doi/pdf/10.1002/pssb.201147052},
  \urlprefix\url{https://onlinelibrary.wiley.com/doi/abs/10.1002/pssb.201147052}.

\bibitem[{\citenamefont{Piefke and Lechermann}(2018)}]{Lechermann_2018}
\bibinfo{author}{\bibfnamefont{C.}~\bibnamefont{Piefke}} \bibnamefont{and}
  \bibinfo{author}{\bibfnamefont{F.}~\bibnamefont{Lechermann}},
  \bibinfo{journal}{Phys. Rev. B} \textbf{\bibinfo{volume}{97}},
  \bibinfo{pages}{125154} (\bibinfo{year}{2018}),
  \urlprefix\url{https://link.aps.org/doi/10.1103/PhysRevB.97.125154}.

\bibitem[{\citenamefont{de' Medici et~al.}(2014)\citenamefont{de' Medici,
  Giovannetti, and Capone}}]{Medici_2014}
\bibinfo{author}{\bibfnamefont{L.}~\bibnamefont{de' Medici}},
  \bibinfo{author}{\bibfnamefont{G.}~\bibnamefont{Giovannetti}},
  \bibnamefont{and} \bibinfo{author}{\bibfnamefont{M.}~\bibnamefont{Capone}},
  \bibinfo{journal}{Phys. Rev. Lett.} \textbf{\bibinfo{volume}{112}},
  \bibinfo{pages}{177001} (\bibinfo{year}{2014}),
  \urlprefix\url{https://link.aps.org/doi/10.1103/PhysRevLett.112.177001}.

\bibitem[{\citenamefont{Frank et~al.}(2021)\citenamefont{Frank, Lee,
  Bhattacharyya, Tsang, Quito, Dobrosavljevi\ifmmode~\acute{c}\else \'{c}\fi{},
  Christiansen, and Lanat\`a}}]{gRISB_2021}
\bibinfo{author}{\bibfnamefont{M.~S.} \bibnamefont{Frank}},
  \bibinfo{author}{\bibfnamefont{T.-H.} \bibnamefont{Lee}},
  \bibinfo{author}{\bibfnamefont{G.}~\bibnamefont{Bhattacharyya}},
  \bibinfo{author}{\bibfnamefont{P.~K.~H.} \bibnamefont{Tsang}},
  \bibinfo{author}{\bibfnamefont{V.~L.} \bibnamefont{Quito}},
  \bibinfo{author}{\bibfnamefont{V.}~\bibnamefont{Dobrosavljevi\ifmmode~\acute{c}\else
  \'{c}\fi{}}}, \bibinfo{author}{\bibfnamefont{O.}~\bibnamefont{Christiansen}},
  \bibnamefont{and} \bibinfo{author}{\bibfnamefont{N.}~\bibnamefont{Lanat\`a}},
  \bibinfo{journal}{Phys. Rev. B} \textbf{\bibinfo{volume}{104}},
  \bibinfo{pages}{L081103} (\bibinfo{year}{2021}),
  \urlprefix\url{https://link.aps.org/doi/10.1103/PhysRevB.104.L081103}.

\bibitem[{\citenamefont{Facio et~al.}(2018)\citenamefont{Facio, Mravlje,
  Pourovskii, Cornaglia, and Vildosola}}]{Facio_2018_PRB}
\bibinfo{author}{\bibfnamefont{J.~I.} \bibnamefont{Facio}},
  \bibinfo{author}{\bibfnamefont{J.}~\bibnamefont{Mravlje}},
  \bibinfo{author}{\bibfnamefont{L.}~\bibnamefont{Pourovskii}},
  \bibinfo{author}{\bibfnamefont{P.~S.} \bibnamefont{Cornaglia}},
  \bibnamefont{and}
  \bibinfo{author}{\bibfnamefont{V.}~\bibnamefont{Vildosola}},
  \bibinfo{journal}{Phys. Rev. B} \textbf{\bibinfo{volume}{98}},
  \bibinfo{pages}{085121} (\bibinfo{year}{2018}),
  \urlprefix\url{https://link.aps.org/doi/10.1103/PhysRevB.98.085121}.

\bibitem[{\citenamefont{Barber et~al.}(2019)\citenamefont{Barber, Lechermann,
  Streltsov, Skornyakov, Ghosh, Ramshaw, Kikugawa, Sokolov, Mackenzie, Hicks
  et~al.}}]{Lechermann_2019_SRO}
\bibinfo{author}{\bibfnamefont{M.~E.} \bibnamefont{Barber}},
  \bibinfo{author}{\bibfnamefont{F.}~\bibnamefont{Lechermann}},
  \bibinfo{author}{\bibfnamefont{S.~V.} \bibnamefont{Streltsov}},
  \bibinfo{author}{\bibfnamefont{S.~L.} \bibnamefont{Skornyakov}},
  \bibinfo{author}{\bibfnamefont{S.}~\bibnamefont{Ghosh}},
  \bibinfo{author}{\bibfnamefont{B.~J.} \bibnamefont{Ramshaw}},
  \bibinfo{author}{\bibfnamefont{N.}~\bibnamefont{Kikugawa}},
  \bibinfo{author}{\bibfnamefont{D.~A.} \bibnamefont{Sokolov}},
  \bibinfo{author}{\bibfnamefont{A.~P.} \bibnamefont{Mackenzie}},
  \bibinfo{author}{\bibfnamefont{C.~W.} \bibnamefont{Hicks}},
  \bibnamefont{et~al.}, \bibinfo{journal}{Phys. Rev. B}
  \textbf{\bibinfo{volume}{100}}, \bibinfo{pages}{245139}
  (\bibinfo{year}{2019}),
  \urlprefix\url{https://link.aps.org/doi/10.1103/PhysRevB.100.245139}.

\bibitem[{\citenamefont{Lanat\`a
  et~al.}(2017{\natexlab{b}})\citenamefont{Lanat\`a, Lee, Yao, and
  Dobrosavljevi\ifmmode~\acute{c}\else \'{c}\fi{}}}]{gRISB_2017}
\bibinfo{author}{\bibfnamefont{N.}~\bibnamefont{Lanat\`a}},
  \bibinfo{author}{\bibfnamefont{T.-H.} \bibnamefont{Lee}},
  \bibinfo{author}{\bibfnamefont{Y.-X.} \bibnamefont{Yao}}, \bibnamefont{and}
  \bibinfo{author}{\bibfnamefont{V.}~\bibnamefont{Dobrosavljevi\ifmmode~\acute{c}\else
  \'{c}\fi{}}}, \bibinfo{journal}{Phys. Rev. B} \textbf{\bibinfo{volume}{96}},
  \bibinfo{pages}{195126} (\bibinfo{year}{2017}{\natexlab{b}}),
  \urlprefix\url{https://link.aps.org/doi/10.1103/PhysRevB.96.195126}.

\bibitem[{\citenamefont{Lanat\`a}(2022)}]{gRISB_2022}
\bibinfo{author}{\bibfnamefont{N.}~\bibnamefont{Lanat\`a}},
  \bibinfo{journal}{Phys. Rev. B} \textbf{\bibinfo{volume}{105}},
  \bibinfo{pages}{045111} (\bibinfo{year}{2022}),
  \urlprefix\url{https://link.aps.org/doi/10.1103/PhysRevB.105.045111}.

\bibitem[{\citenamefont{Fertitta and Booth}(2018)}]{Fertitta2018}
\bibinfo{author}{\bibfnamefont{E.}~\bibnamefont{Fertitta}} \bibnamefont{and}
  \bibinfo{author}{\bibfnamefont{G.~H.} \bibnamefont{Booth}},
  \bibinfo{journal}{Phys. Rev. B} \textbf{\bibinfo{volume}{98}},
  \bibinfo{pages}{235132} (\bibinfo{year}{2018}),
  \urlprefix\url{https://link.aps.org/doi/10.1103/PhysRevB.98.235132}.

\bibitem[{\citenamefont{Sriluckshmy et~al.}(2021)\citenamefont{Sriluckshmy,
  Nusspickel, Fertitta, and Booth}}]{Sriluckshmy2021}
\bibinfo{author}{\bibfnamefont{P.~V.} \bibnamefont{Sriluckshmy}},
  \bibinfo{author}{\bibfnamefont{M.}~\bibnamefont{Nusspickel}},
  \bibinfo{author}{\bibfnamefont{E.}~\bibnamefont{Fertitta}}, \bibnamefont{and}
  \bibinfo{author}{\bibfnamefont{G.~H.} \bibnamefont{Booth}},
  \bibinfo{journal}{Phys. Rev. B} \textbf{\bibinfo{volume}{103}},
  \bibinfo{pages}{085131} (\bibinfo{year}{2021}),
  \urlprefix\url{https://link.aps.org/doi/10.1103/PhysRevB.103.085131}.

\bibitem[{\citenamefont{Zhang and Sachdev}(2020)}]{ancilla_quibits_2020}
\bibinfo{author}{\bibfnamefont{Y.-H.} \bibnamefont{Zhang}} \bibnamefont{and}
  \bibinfo{author}{\bibfnamefont{S.}~\bibnamefont{Sachdev}},
  \bibinfo{journal}{Phys. Rev. Res.} \textbf{\bibinfo{volume}{2}},
  \bibinfo{pages}{023172} (\bibinfo{year}{2020}),
  \urlprefix\url{https://link.aps.org/doi/10.1103/PhysRevResearch.2.023172}.

\bibitem[{\citenamefont{Moreno et~al.}(2022)\citenamefont{Moreno, Carleo,
  Georges, and Stokes}}]{hidden_fermion}
\bibinfo{author}{\bibfnamefont{J.~R.} \bibnamefont{Moreno}},
  \bibinfo{author}{\bibfnamefont{G.}~\bibnamefont{Carleo}},
  \bibinfo{author}{\bibfnamefont{A.}~\bibnamefont{Georges}}, \bibnamefont{and}
  \bibinfo{author}{\bibfnamefont{J.}~\bibnamefont{Stokes}},
  \bibinfo{journal}{Proceedings of the National Academy of Sciences}
  \textbf{\bibinfo{volume}{119}}, \bibinfo{pages}{e2122059119}
  (\bibinfo{year}{2022}),
  \eprint{https://www.pnas.org/doi/pdf/10.1073/pnas.2122059119},
  \urlprefix\url{https://www.pnas.org/doi/abs/10.1073/pnas.2122059119}.

\bibitem[{\citenamefont{Guerci et~al.}(2019)\citenamefont{Guerci, Capone, and
  Fabrizio}}]{Guerci2019}
\bibinfo{author}{\bibfnamefont{D.}~\bibnamefont{Guerci}},
  \bibinfo{author}{\bibfnamefont{M.}~\bibnamefont{Capone}}, \bibnamefont{and}
  \bibinfo{author}{\bibfnamefont{M.}~\bibnamefont{Fabrizio}},
  \bibinfo{journal}{Phys. Rev. Materials} \textbf{\bibinfo{volume}{3}},
  \bibinfo{pages}{054605} (\bibinfo{year}{2019}),
  \urlprefix\url{https://link.aps.org/doi/10.1103/PhysRevMaterials.3.054605}.

\bibitem[{\citenamefont{Lee et~al.}(2023)\citenamefont{Lee, Lanat\`a, and
  Kotliar}}]{gRISB_accuracy}
\bibinfo{author}{\bibfnamefont{T.-H.} \bibnamefont{Lee}},
  \bibinfo{author}{\bibfnamefont{N.}~\bibnamefont{Lanat\`a}}, \bibnamefont{and}
  \bibinfo{author}{\bibfnamefont{G.}~\bibnamefont{Kotliar}},
  \bibinfo{journal}{Phys. Rev. B} \textbf{\bibinfo{volume}{107}},
  \bibinfo{pages}{L121104} (\bibinfo{year}{2023}),
  \urlprefix\url{https://link.aps.org/doi/10.1103/PhysRevB.107.L121104}.

\bibitem[{\citenamefont{Guerci}(2019)}]{Guerci_thesis}
\bibinfo{author}{\bibfnamefont{D.}~\bibnamefont{Guerci}}, Ph.D. thesis,
  \bibinfo{school}{International School for Advanced Studies},
  \bibinfo{address}{https://iris.sissa.it/handle/20.500.11767/103994}
  (\bibinfo{year}{2019}).

\bibitem[{\citenamefont{Guerci et~al.}(2023)\citenamefont{Guerci, Capone, and
  Lanata}}]{Guerci_TDgGA}
\bibinfo{author}{\bibfnamefont{D.}~\bibnamefont{Guerci}},
  \bibinfo{author}{\bibfnamefont{M.}~\bibnamefont{Capone}}, \bibnamefont{and}
  \bibinfo{author}{\bibfnamefont{N.}~\bibnamefont{Lanata}},
  \bibinfo{journal}{arXiv:2303.09584}  (\bibinfo{year}{2023}),
  \urlprefix\url{https://doi.org/10.48550/arXiv.2303.09584}.

\bibitem[{\citenamefont{Mejuto-Zaera and Fabrizio}(2023)}]{Carlos2023}
\bibinfo{author}{\bibfnamefont{C.}~\bibnamefont{Mejuto-Zaera}}
  \bibnamefont{and} \bibinfo{author}{\bibfnamefont{M.}~\bibnamefont{Fabrizio}},
  \bibinfo{journal}{Phys. Rev. B} \textbf{\bibinfo{volume}{107}},
  \bibinfo{pages}{235150} (\bibinfo{year}{2023}),
  \urlprefix\url{https://link.aps.org/doi/10.1103/PhysRevB.107.235150}.

\bibitem[{\citenamefont{White}(1992)}]{White1992}
\bibinfo{author}{\bibfnamefont{S.~R.} \bibnamefont{White}},
  \bibinfo{journal}{Phys. Rev. Lett.} \textbf{\bibinfo{volume}{69}},
  \bibinfo{pages}{2863} (\bibinfo{year}{1992}),
  \urlprefix\url{https://link.aps.org/doi/10.1103/PhysRevLett.69.2863}.

\bibitem[{\citenamefont{White}(1993)}]{White1993}
\bibinfo{author}{\bibfnamefont{S.~R.} \bibnamefont{White}},
  \bibinfo{journal}{Phys. Rev. B} \textbf{\bibinfo{volume}{48}},
  \bibinfo{pages}{10345} (\bibinfo{year}{1993}),
  \urlprefix\url{https://link.aps.org/doi/10.1103/PhysRevB.48.10345}.

\bibitem[{\citenamefont{Schollw\"ock}(2005)}]{Schollwock2005}
\bibinfo{author}{\bibfnamefont{U.}~\bibnamefont{Schollw\"ock}},
  \bibinfo{journal}{Rev. Mod. Phys.} \textbf{\bibinfo{volume}{77}},
  \bibinfo{pages}{259} (\bibinfo{year}{2005}),
  \urlprefix\url{https://link.aps.org/doi/10.1103/RevModPhys.77.259}.

\bibitem[{\citenamefont{Kanamori}(1963)}]{Kanamori_1963}
\bibinfo{author}{\bibfnamefont{J.}~\bibnamefont{Kanamori}},
  \bibinfo{journal}{Progress of Theoretical Physics}
  \textbf{\bibinfo{volume}{30}}, \bibinfo{pages}{275} (\bibinfo{year}{1963}),
  ISSN \bibinfo{issn}{0033-068X},
  \eprint{https://academic.oup.com/ptp/article-pdf/30/3/275/5278869/30-3-275.pdf},
  \urlprefix\url{https://doi.org/10.1143/PTP.30.275}.

\bibitem[{\citenamefont{Ayral et~al.}(2017)\citenamefont{Ayral, Lee, and
  Kotliar}}]{RISB_DMET_Ayral_2018}
\bibinfo{author}{\bibfnamefont{T.}~\bibnamefont{Ayral}},
  \bibinfo{author}{\bibfnamefont{T.-H.} \bibnamefont{Lee}}, \bibnamefont{and}
  \bibinfo{author}{\bibfnamefont{G.}~\bibnamefont{Kotliar}},
  \bibinfo{journal}{Phys. Rev. B} \textbf{\bibinfo{volume}{96}},
  \bibinfo{pages}{235139} (\bibinfo{year}{2017}),
  \urlprefix\url{https://link.aps.org/doi/10.1103/PhysRevB.96.235139}.

\bibitem[{\citenamefont{Lanat\`a}(2023)}]{gDMET}
\bibinfo{author}{\bibfnamefont{N.}~\bibnamefont{Lanat\`a}},
  \bibinfo{journal}{Phys. Rev. B} \textbf{\bibinfo{volume}{108}},
  \bibinfo{pages}{235112} (\bibinfo{year}{2023}),
  \urlprefix\url{https://link.aps.org/doi/10.1103/PhysRevB.108.235112}.

\bibitem[{\citenamefont{Sun et~al.}(2018)\citenamefont{Sun, Berkelbach, Blunt,
  Booth, Guo, Li, Liu, McClain, Sayfutyarova, Sharma et~al.}}]{PySCF_2018}
\bibinfo{author}{\bibfnamefont{Q.}~\bibnamefont{Sun}},
  \bibinfo{author}{\bibfnamefont{T.~C.} \bibnamefont{Berkelbach}},
  \bibinfo{author}{\bibfnamefont{N.~S.} \bibnamefont{Blunt}},
  \bibinfo{author}{\bibfnamefont{G.~H.} \bibnamefont{Booth}},
  \bibinfo{author}{\bibfnamefont{S.}~\bibnamefont{Guo}},
  \bibinfo{author}{\bibfnamefont{Z.}~\bibnamefont{Li}},
  \bibinfo{author}{\bibfnamefont{J.}~\bibnamefont{Liu}},
  \bibinfo{author}{\bibfnamefont{J.~D.} \bibnamefont{McClain}},
  \bibinfo{author}{\bibfnamefont{E.~R.} \bibnamefont{Sayfutyarova}},
  \bibinfo{author}{\bibfnamefont{S.}~\bibnamefont{Sharma}},
  \bibnamefont{et~al.}, \bibinfo{journal}{WIREs Computational Molecular
  Science} \textbf{\bibinfo{volume}{8}}, \bibinfo{pages}{e1340}
  (\bibinfo{year}{2018}),
  \urlprefix\url{https://wires.onlinelibrary.wiley.com/doi/abs/10.1002/wcms.1340}.

\bibitem[{\citenamefont{Sun et~al.}(2020)\citenamefont{Sun, Zhang, Banerjee,
  Bao, Barbry, Blunt, Bogdanov, Booth, Chen, Cui et~al.}}]{PySCF_2020}
\bibinfo{author}{\bibfnamefont{Q.}~\bibnamefont{Sun}},
  \bibinfo{author}{\bibfnamefont{X.}~\bibnamefont{Zhang}},
  \bibinfo{author}{\bibfnamefont{S.}~\bibnamefont{Banerjee}},
  \bibinfo{author}{\bibfnamefont{P.}~\bibnamefont{Bao}},
  \bibinfo{author}{\bibfnamefont{M.}~\bibnamefont{Barbry}},
  \bibinfo{author}{\bibfnamefont{N.~S.} \bibnamefont{Blunt}},
  \bibinfo{author}{\bibfnamefont{N.~A.} \bibnamefont{Bogdanov}},
  \bibinfo{author}{\bibfnamefont{G.~H.} \bibnamefont{Booth}},
  \bibinfo{author}{\bibfnamefont{J.}~\bibnamefont{Chen}},
  \bibinfo{author}{\bibfnamefont{Z.-H.} \bibnamefont{Cui}},
  \bibnamefont{et~al.}, \bibinfo{journal}{The Journal of Chemical Physics}
  \textbf{\bibinfo{volume}{153}}, \bibinfo{pages}{024109}
  (\bibinfo{year}{2020}), \eprint{https://doi.org/10.1063/5.0006074},
  \urlprefix\url{https://doi.org/10.1063/5.0006074}.

\bibitem[{\citenamefont{Zhai and Chan}(2021)}]{block2}
\bibinfo{author}{\bibfnamefont{H.}~\bibnamefont{Zhai}} \bibnamefont{and}
  \bibinfo{author}{\bibfnamefont{G.~K.-L.} \bibnamefont{Chan}},
  \bibinfo{journal}{The Journal of Chemical Physics}
  \textbf{\bibinfo{volume}{154}}, \bibinfo{pages}{224116}
  (\bibinfo{year}{2021}), \eprint{https://doi.org/10.1063/5.0050902},
  \urlprefix\url{https://doi.org/10.1063/5.0050902}.

\bibitem[{\citenamefont{Zhai et~al.}(2023)\citenamefont{Zhai, Larsson, Lee,
  Zhi-Hao, Zhu, Sung, Peng, Peng, Liao, T\"{o}lle et~al.}}]{block2_2}
\bibinfo{author}{\bibfnamefont{H.}~\bibnamefont{Zhai}},
  \bibinfo{author}{\bibfnamefont{H.~R.} \bibnamefont{Larsson}},
  \bibinfo{author}{\bibfnamefont{S.~L.} \bibnamefont{Lee}},
  \bibinfo{author}{\bibfnamefont{C.}~\bibnamefont{Zhi-Hao}},
  \bibinfo{author}{\bibfnamefont{T.}~\bibnamefont{Zhu}},
  \bibinfo{author}{\bibfnamefont{C.}~\bibnamefont{Sung}},
  \bibinfo{author}{\bibfnamefont{L.}~\bibnamefont{Peng}},
  \bibinfo{author}{\bibfnamefont{R.}~\bibnamefont{Peng}},
  \bibinfo{author}{\bibfnamefont{K.}~\bibnamefont{Liao}},
  \bibinfo{author}{\bibfnamefont{J.}~\bibnamefont{T\"{o}lle}},
  \bibnamefont{et~al.}, \bibinfo{journal}{arXiv:2310.03920}
  (\bibinfo{year}{2023}), \urlprefix\url{https://arxiv.org/abs/2310.03920}.

\bibitem[{\citenamefont{Savrasov et~al.}(2006)\citenamefont{Savrasov, Haule,
  and Kotliar}}]{Savrasov2006}
\bibinfo{author}{\bibfnamefont{S.~Y.} \bibnamefont{Savrasov}},
  \bibinfo{author}{\bibfnamefont{K.}~\bibnamefont{Haule}}, \bibnamefont{and}
  \bibinfo{author}{\bibfnamefont{G.}~\bibnamefont{Kotliar}},
  \bibinfo{journal}{Phys. Rev. Lett.} \textbf{\bibinfo{volume}{96}},
  \bibinfo{pages}{036404} (\bibinfo{year}{2006}),
  \urlprefix\url{https://link.aps.org/doi/10.1103/PhysRevLett.96.036404}.

\bibitem[{\citenamefont{Caffarel and Krauth}(1994)}]{Caffarel_Krauth_1994}
\bibinfo{author}{\bibfnamefont{M.}~\bibnamefont{Caffarel}} \bibnamefont{and}
  \bibinfo{author}{\bibfnamefont{W.}~\bibnamefont{Krauth}},
  \bibinfo{journal}{Phys. Rev. Lett.} \textbf{\bibinfo{volume}{72}},
  \bibinfo{pages}{1545} (\bibinfo{year}{1994}),
  \urlprefix\url{https://link.aps.org/doi/10.1103/PhysRevLett.72.1545}.

\bibitem[{\citenamefont{Rozenberg et~al.}(1994)\citenamefont{Rozenberg,
  Moeller, and Kotliar}}]{Rozenberg1994}
\bibinfo{author}{\bibfnamefont{M.~J.} \bibnamefont{Rozenberg}},
  \bibinfo{author}{\bibfnamefont{G.}~\bibnamefont{Moeller}}, \bibnamefont{and}
  \bibinfo{author}{\bibfnamefont{G.}~\bibnamefont{Kotliar}},
  \bibinfo{journal}{Modern Physics Letters B} \textbf{\bibinfo{volume}{08}},
  \bibinfo{pages}{535} (\bibinfo{year}{1994}),
  \eprint{https://doi.org/10.1142/S0217984994000571},
  \urlprefix\url{https://doi.org/10.1142/S0217984994000571}.

\bibitem[{\citenamefont{Liebsch}(2005)}]{Liebsch_2005}
\bibinfo{author}{\bibfnamefont{A.}~\bibnamefont{Liebsch}},
  \bibinfo{journal}{Phys. Rev. Lett.} \textbf{\bibinfo{volume}{95}},
  \bibinfo{pages}{116402} (\bibinfo{year}{2005}),
  \urlprefix\url{https://link.aps.org/doi/10.1103/PhysRevLett.95.116402}.

\bibitem[{\citenamefont{Perroni et~al.}(2007)\citenamefont{Perroni, Ishida, and
  Liebsch}}]{Liebsch_2007}
\bibinfo{author}{\bibfnamefont{C.~A.} \bibnamefont{Perroni}},
  \bibinfo{author}{\bibfnamefont{H.}~\bibnamefont{Ishida}}, \bibnamefont{and}
  \bibinfo{author}{\bibfnamefont{A.}~\bibnamefont{Liebsch}},
  \bibinfo{journal}{Phys. Rev. B} \textbf{\bibinfo{volume}{75}},
  \bibinfo{pages}{045125} (\bibinfo{year}{2007}),
  \urlprefix\url{https://link.aps.org/doi/10.1103/PhysRevB.75.045125}.

\bibitem[{\citenamefont{Ishida and Liebsch}(2010)}]{Liebsch_2010}
\bibinfo{author}{\bibfnamefont{H.}~\bibnamefont{Ishida}} \bibnamefont{and}
  \bibinfo{author}{\bibfnamefont{A.}~\bibnamefont{Liebsch}},
  \bibinfo{journal}{Phys. Rev. B} \textbf{\bibinfo{volume}{81}},
  \bibinfo{pages}{054513} (\bibinfo{year}{2010}),
  \urlprefix\url{https://link.aps.org/doi/10.1103/PhysRevB.81.054513}.

\bibitem[{\citenamefont{Liebsch and Ishida}(2011)}]{Liebsch_2011}
\bibinfo{author}{\bibfnamefont{A.}~\bibnamefont{Liebsch}} \bibnamefont{and}
  \bibinfo{author}{\bibfnamefont{H.}~\bibnamefont{Ishida}},
  \bibinfo{journal}{Journal of Physics: Condensed Matter}
  \textbf{\bibinfo{volume}{24}}, \bibinfo{pages}{053201}
  (\bibinfo{year}{2011}),
  \urlprefix\url{https://doi.org/10.1088/0953-8984/24/5/053201}.

\bibitem[{\citenamefont{Capone et~al.}(2004)\citenamefont{Capone, Civelli,
  Kancharla, Castellani, and Kotliar}}]{Capone2004}
\bibinfo{author}{\bibfnamefont{M.}~\bibnamefont{Capone}},
  \bibinfo{author}{\bibfnamefont{M.}~\bibnamefont{Civelli}},
  \bibinfo{author}{\bibfnamefont{S.~S.} \bibnamefont{Kancharla}},
  \bibinfo{author}{\bibfnamefont{C.}~\bibnamefont{Castellani}},
  \bibnamefont{and} \bibinfo{author}{\bibfnamefont{G.}~\bibnamefont{Kotliar}},
  \bibinfo{journal}{Phys. Rev. B} \textbf{\bibinfo{volume}{69}},
  \bibinfo{pages}{195105} (\bibinfo{year}{2004}),
  \urlprefix\url{https://link.aps.org/doi/10.1103/PhysRevB.69.195105}.

\bibitem[{\citenamefont{Civelli et~al.}(2005)\citenamefont{Civelli, Capone,
  Kancharla, Parcollet, and Kotliar}}]{Civelli_2005}
\bibinfo{author}{\bibfnamefont{M.}~\bibnamefont{Civelli}},
  \bibinfo{author}{\bibfnamefont{M.}~\bibnamefont{Capone}},
  \bibinfo{author}{\bibfnamefont{S.~S.} \bibnamefont{Kancharla}},
  \bibinfo{author}{\bibfnamefont{O.}~\bibnamefont{Parcollet}},
  \bibnamefont{and} \bibinfo{author}{\bibfnamefont{G.}~\bibnamefont{Kotliar}},
  \bibinfo{journal}{Phys. Rev. Lett.} \textbf{\bibinfo{volume}{95}},
  \bibinfo{pages}{106402} (\bibinfo{year}{2005}),
  \urlprefix\url{https://link.aps.org/doi/10.1103/PhysRevLett.95.106402}.

\bibitem[{\citenamefont{Kyung et~al.}(2006)\citenamefont{Kyung, Kancharla,
  S\'en\'echal, Tremblay, Civelli, and Kotliar}}]{Kyung_2006}
\bibinfo{author}{\bibfnamefont{B.}~\bibnamefont{Kyung}},
  \bibinfo{author}{\bibfnamefont{S.~S.} \bibnamefont{Kancharla}},
  \bibinfo{author}{\bibfnamefont{D.}~\bibnamefont{S\'en\'echal}},
  \bibinfo{author}{\bibfnamefont{A.-M.~S.} \bibnamefont{Tremblay}},
  \bibinfo{author}{\bibfnamefont{M.}~\bibnamefont{Civelli}}, \bibnamefont{and}
  \bibinfo{author}{\bibfnamefont{G.}~\bibnamefont{Kotliar}},
  \bibinfo{journal}{Phys. Rev. B} \textbf{\bibinfo{volume}{73}},
  \bibinfo{pages}{165114} (\bibinfo{year}{2006}),
  \urlprefix\url{https://link.aps.org/doi/10.1103/PhysRevB.73.165114}.

\bibitem[{\citenamefont{Amaricci et~al.}(2022)\citenamefont{Amaricci, Crippa,
  Scazzola, Petocchi, Mazza, {de Medici}, and Capone}}]{AMARICCI_EDI}
\bibinfo{author}{\bibfnamefont{A.}~\bibnamefont{Amaricci}},
  \bibinfo{author}{\bibfnamefont{L.}~\bibnamefont{Crippa}},
  \bibinfo{author}{\bibfnamefont{A.}~\bibnamefont{Scazzola}},
  \bibinfo{author}{\bibfnamefont{F.}~\bibnamefont{Petocchi}},
  \bibinfo{author}{\bibfnamefont{G.}~\bibnamefont{Mazza}},
  \bibinfo{author}{\bibfnamefont{L.}~\bibnamefont{{de Medici}}},
  \bibnamefont{and} \bibinfo{author}{\bibfnamefont{M.}~\bibnamefont{Capone}},
  \bibinfo{journal}{Computer Physics Communications}
  \textbf{\bibinfo{volume}{273}}, \bibinfo{pages}{108261}
  (\bibinfo{year}{2022}), ISSN \bibinfo{issn}{0010-4655},
  \urlprefix\url{https://www.sciencedirect.com/science/article/pii/S0010465521003738}.

\bibitem[{\citenamefont{Iskakov and Danilov}(2018)}]{ISKAKOV2018128}
\bibinfo{author}{\bibfnamefont{S.}~\bibnamefont{Iskakov}} \bibnamefont{and}
  \bibinfo{author}{\bibfnamefont{M.}~\bibnamefont{Danilov}},
  \bibinfo{journal}{Computer Physics Communications}
  \textbf{\bibinfo{volume}{225}}, \bibinfo{pages}{128} (\bibinfo{year}{2018}),
  ISSN \bibinfo{issn}{0010-4655},
  \urlprefix\url{https://www.sciencedirect.com/science/article/pii/S0010465517304216}.

\bibitem[{\citenamefont{Capone et~al.}(2007)\citenamefont{Capone, de' Medici,
  and Georges}}]{Capone2007}
\bibinfo{author}{\bibfnamefont{M.}~\bibnamefont{Capone}},
  \bibinfo{author}{\bibfnamefont{L.}~\bibnamefont{de' Medici}},
  \bibnamefont{and} \bibinfo{author}{\bibfnamefont{A.}~\bibnamefont{Georges}},
  \bibinfo{journal}{Phys. Rev. B} \textbf{\bibinfo{volume}{76}},
  \bibinfo{pages}{245116} (\bibinfo{year}{2007}),
  \urlprefix\url{https://link.aps.org/doi/10.1103/PhysRevB.76.245116}.

\bibitem[{\citenamefont{Parcollet et~al.}(2015)\citenamefont{Parcollet,
  Ferrero, Ayral, Hafermann, Krivenko, Messio, and Seth}}]{triqs}
\bibinfo{author}{\bibfnamefont{O.}~\bibnamefont{Parcollet}},
  \bibinfo{author}{\bibfnamefont{M.}~\bibnamefont{Ferrero}},
  \bibinfo{author}{\bibfnamefont{T.}~\bibnamefont{Ayral}},
  \bibinfo{author}{\bibfnamefont{H.}~\bibnamefont{Hafermann}},
  \bibinfo{author}{\bibfnamefont{I.}~\bibnamefont{Krivenko}},
  \bibinfo{author}{\bibfnamefont{L.}~\bibnamefont{Messio}}, \bibnamefont{and}
  \bibinfo{author}{\bibfnamefont{P.}~\bibnamefont{Seth}},
  \bibinfo{journal}{Computer Physics Communications}
  \textbf{\bibinfo{volume}{196}}, \bibinfo{pages}{398 } (\bibinfo{year}{2015}),
  ISSN \bibinfo{issn}{0010-4655},
  \urlprefix\url{http://www.sciencedirect.com/science/article/pii/S0010465515001666}.

\bibitem[{\citenamefont{Seth et~al.}(2016)\citenamefont{Seth, Krivenko,
  Ferrero, and Parcollet}}]{TRIQS_CTHYB}
\bibinfo{author}{\bibfnamefont{P.}~\bibnamefont{Seth}},
  \bibinfo{author}{\bibfnamefont{I.}~\bibnamefont{Krivenko}},
  \bibinfo{author}{\bibfnamefont{M.}~\bibnamefont{Ferrero}}, \bibnamefont{and}
  \bibinfo{author}{\bibfnamefont{O.}~\bibnamefont{Parcollet}},
  \bibinfo{journal}{Computer Physics Communications}
  \textbf{\bibinfo{volume}{200}}, \bibinfo{pages}{274 } (\bibinfo{year}{2016}),
  ISSN \bibinfo{issn}{0010-4655},
  \urlprefix\url{http://www.sciencedirect.com/science/article/pii/S001046551500404X}.

\bibitem[{\citenamefont{Gull et~al.}(2011)\citenamefont{Gull, Millis,
  Lichtenstein, Rubtsov, Troyer, and Werner}}]{Gull_RMP_2011}
\bibinfo{author}{\bibfnamefont{E.}~\bibnamefont{Gull}},
  \bibinfo{author}{\bibfnamefont{A.~J.} \bibnamefont{Millis}},
  \bibinfo{author}{\bibfnamefont{A.~I.} \bibnamefont{Lichtenstein}},
  \bibinfo{author}{\bibfnamefont{A.~N.} \bibnamefont{Rubtsov}},
  \bibinfo{author}{\bibfnamefont{M.}~\bibnamefont{Troyer}}, \bibnamefont{and}
  \bibinfo{author}{\bibfnamefont{P.}~\bibnamefont{Werner}},
  \bibinfo{journal}{Rev. Mod. Phys.} \textbf{\bibinfo{volume}{83}},
  \bibinfo{pages}{349} (\bibinfo{year}{2011}),
  \urlprefix\url{https://link.aps.org/doi/10.1103/RevModPhys.83.349}.

\bibitem[{\citenamefont{Melnick et~al.}(2021)\citenamefont{Melnick, S\'emon,
  Yu, D'Imperio, Tremblay, and Kotliar}}]{MELNICK2021108075}
\bibinfo{author}{\bibfnamefont{C.}~\bibnamefont{Melnick}},
  \bibinfo{author}{\bibfnamefont{P.}~\bibnamefont{S\'emon}},
  \bibinfo{author}{\bibfnamefont{K.}~\bibnamefont{Yu}},
  \bibinfo{author}{\bibfnamefont{N.}~\bibnamefont{D'Imperio}},
  \bibinfo{author}{\bibfnamefont{A.-M.} \bibnamefont{Tremblay}},
  \bibnamefont{and} \bibinfo{author}{\bibfnamefont{G.}~\bibnamefont{Kotliar}},
  \bibinfo{journal}{Computer Physics Communications}
  \textbf{\bibinfo{volume}{267}}, \bibinfo{pages}{108075}
  (\bibinfo{year}{2021}), ISSN \bibinfo{issn}{0010-4655},
  \urlprefix\url{https://www.sciencedirect.com/science/article/pii/S0010465521001879}.

\bibitem[{\citenamefont{Chatzieleftheriou
  et~al.}(2023)\citenamefont{Chatzieleftheriou, Kowalski,
  Berovi\ifmmode~\acute{c}\else \'{c}\fi{}, Amaricci, Capone, De~Leo,
  Sangiovanni, and de' Medici}}]{DeMedici2022}
\bibinfo{author}{\bibfnamefont{M.}~\bibnamefont{Chatzieleftheriou}},
  \bibinfo{author}{\bibfnamefont{A.}~\bibnamefont{Kowalski}},
  \bibinfo{author}{\bibfnamefont{M.}~\bibnamefont{Berovi\ifmmode~\acute{c}\else
  \'{c}\fi{}}}, \bibinfo{author}{\bibfnamefont{A.}~\bibnamefont{Amaricci}},
  \bibinfo{author}{\bibfnamefont{M.}~\bibnamefont{Capone}},
  \bibinfo{author}{\bibfnamefont{L.}~\bibnamefont{De~Leo}},
  \bibinfo{author}{\bibfnamefont{G.}~\bibnamefont{Sangiovanni}},
  \bibnamefont{and} \bibinfo{author}{\bibfnamefont{L.}~\bibnamefont{de'
  Medici}}, \bibinfo{journal}{Phys. Rev. Lett.} \textbf{\bibinfo{volume}{130}},
  \bibinfo{pages}{066401} (\bibinfo{year}{2023}),
  \urlprefix\url{https://link.aps.org/doi/10.1103/PhysRevLett.130.066401}.

\bibitem[{\citenamefont{Tamai et~al.}(2019)\citenamefont{Tamai, Zingl,
  Rozbicki, Cappelli, Ricc\`o, de~la Torre, McKeown~Walker, Bruno, King,
  Meevasana et~al.}}]{Tamai_Zingl_Georges_2019_PRX}
\bibinfo{author}{\bibfnamefont{A.}~\bibnamefont{Tamai}},
  \bibinfo{author}{\bibfnamefont{M.}~\bibnamefont{Zingl}},
  \bibinfo{author}{\bibfnamefont{E.}~\bibnamefont{Rozbicki}},
  \bibinfo{author}{\bibfnamefont{E.}~\bibnamefont{Cappelli}},
  \bibinfo{author}{\bibfnamefont{S.}~\bibnamefont{Ricc\`o}},
  \bibinfo{author}{\bibfnamefont{A.}~\bibnamefont{de~la Torre}},
  \bibinfo{author}{\bibfnamefont{S.}~\bibnamefont{McKeown~Walker}},
  \bibinfo{author}{\bibfnamefont{F.~Y.} \bibnamefont{Bruno}},
  \bibinfo{author}{\bibfnamefont{P.~D.~C.} \bibnamefont{King}},
  \bibinfo{author}{\bibfnamefont{W.}~\bibnamefont{Meevasana}},
  \bibnamefont{et~al.}, \bibinfo{journal}{Phys. Rev. X}
  \textbf{\bibinfo{volume}{9}}, \bibinfo{pages}{021048} (\bibinfo{year}{2019}),
  \urlprefix\url{https://link.aps.org/doi/10.1103/PhysRevX.9.021048}.

\bibitem[{\citenamefont{Deng et~al.}(2016)\citenamefont{Deng, Haule, and
  Kotliar}}]{Xiaoyu_Ru_PRL}
\bibinfo{author}{\bibfnamefont{X.}~\bibnamefont{Deng}},
  \bibinfo{author}{\bibfnamefont{K.}~\bibnamefont{Haule}}, \bibnamefont{and}
  \bibinfo{author}{\bibfnamefont{G.}~\bibnamefont{Kotliar}},
  \bibinfo{journal}{Phys. Rev. Lett.} \textbf{\bibinfo{volume}{116}},
  \bibinfo{pages}{256401} (\bibinfo{year}{2016}),
  \urlprefix\url{https://link.aps.org/doi/10.1103/PhysRevLett.116.256401}.

\bibitem[{\citenamefont{Bergemann et~al.}(2003)\citenamefont{Bergemann,
  Mackenzie, Julian, Forsythe, and Ohmichi}}]{Mackenzie_rev_2003}
\bibinfo{author}{\bibfnamefont{C.}~\bibnamefont{Bergemann}},
  \bibinfo{author}{\bibfnamefont{A.~P.} \bibnamefont{Mackenzie}},
  \bibinfo{author}{\bibfnamefont{S.~R.} \bibnamefont{Julian}},
  \bibinfo{author}{\bibfnamefont{D.}~\bibnamefont{Forsythe}}, \bibnamefont{and}
  \bibinfo{author}{\bibfnamefont{E.}~\bibnamefont{Ohmichi}},
  \bibinfo{journal}{Advances in Physics} \textbf{\bibinfo{volume}{52}},
  \bibinfo{pages}{639} (\bibinfo{year}{2003}),
  \eprint{https://doi.org/10.1080/00018730310001621737},
  \urlprefix\url{https://doi.org/10.1080/00018730310001621737}.

\bibitem[{\citenamefont{Hohenberg and Kohn}(1964)}]{Hohenberg1964}
\bibinfo{author}{\bibfnamefont{P.}~\bibnamefont{Hohenberg}} \bibnamefont{and}
  \bibinfo{author}{\bibfnamefont{W.}~\bibnamefont{Kohn}},
  \bibinfo{journal}{Phys. Rev.} \textbf{\bibinfo{volume}{136}},
  \bibinfo{pages}{B864} (\bibinfo{year}{1964}),
  \urlprefix\url{https://link.aps.org/doi/10.1103/PhysRev.136.B864}.

\bibitem[{\citenamefont{Kohn and Sham}(1965)}]{Kohn1965}
\bibinfo{author}{\bibfnamefont{W.}~\bibnamefont{Kohn}} \bibnamefont{and}
  \bibinfo{author}{\bibfnamefont{L.~J.} \bibnamefont{Sham}},
  \bibinfo{journal}{Phys. Rev.} \textbf{\bibinfo{volume}{140}},
  \bibinfo{pages}{A1133} (\bibinfo{year}{1965}),
  \urlprefix\url{https://link.aps.org/doi/10.1103/PhysRev.140.A1133}.

\bibitem[{\citenamefont{Marzari et~al.}(2012)\citenamefont{Marzari, Mostofi,
  Yates, Souza, and Vanderbilt}}]{Mazari2012}
\bibinfo{author}{\bibfnamefont{N.}~\bibnamefont{Marzari}},
  \bibinfo{author}{\bibfnamefont{A.~A.} \bibnamefont{Mostofi}},
  \bibinfo{author}{\bibfnamefont{J.~R.} \bibnamefont{Yates}},
  \bibinfo{author}{\bibfnamefont{I.}~\bibnamefont{Souza}}, \bibnamefont{and}
  \bibinfo{author}{\bibfnamefont{D.}~\bibnamefont{Vanderbilt}},
  \bibinfo{journal}{Rev. Mod. Phys.} \textbf{\bibinfo{volume}{84}},
  \bibinfo{pages}{1419} (\bibinfo{year}{2012}),
  \urlprefix\url{https://link.aps.org/doi/10.1103/RevModPhys.84.1419}.

\bibitem[{\citenamefont{Blaha et~al.}(2020)\citenamefont{Blaha, Schwarz, Tran,
  Laskowski, Madsen, and Marks}}]{Wien2k}
\bibinfo{author}{\bibfnamefont{P.}~\bibnamefont{Blaha}},
  \bibinfo{author}{\bibfnamefont{K.}~\bibnamefont{Schwarz}},
  \bibinfo{author}{\bibfnamefont{F.}~\bibnamefont{Tran}},
  \bibinfo{author}{\bibfnamefont{R.}~\bibnamefont{Laskowski}},
  \bibinfo{author}{\bibfnamefont{G.~K.~H.} \bibnamefont{Madsen}},
  \bibnamefont{and} \bibinfo{author}{\bibfnamefont{L.~D.} \bibnamefont{Marks}},
  \bibinfo{journal}{The Journal of Chemical Physics}
  \textbf{\bibinfo{volume}{152}}, \bibinfo{pages}{074101}
  (\bibinfo{year}{2020}), \eprint{https://doi.org/10.1063/1.5143061},
  \urlprefix\url{https://doi.org/10.1063/1.5143061}.

\bibitem[{\citenamefont{Vogt and Buttrey}(1995)}]{SRO_struct}
\bibinfo{author}{\bibfnamefont{T.}~\bibnamefont{Vogt}} \bibnamefont{and}
  \bibinfo{author}{\bibfnamefont{D.~J.} \bibnamefont{Buttrey}},
  \bibinfo{journal}{Phys. Rev. B} \textbf{\bibinfo{volume}{52}},
  \bibinfo{pages}{R9843} (\bibinfo{year}{1995}),
  \urlprefix\url{https://link.aps.org/doi/10.1103/PhysRevB.52.R9843}.

\bibitem[{\citenamefont{Bl\"ochl et~al.}(1994)\citenamefont{Bl\"ochl, Jepsen,
  and Andersen}}]{tetrahedron}
\bibinfo{author}{\bibfnamefont{P.~E.} \bibnamefont{Bl\"ochl}},
  \bibinfo{author}{\bibfnamefont{O.}~\bibnamefont{Jepsen}}, \bibnamefont{and}
  \bibinfo{author}{\bibfnamefont{O.~K.} \bibnamefont{Andersen}},
  \bibinfo{journal}{Phys. Rev. B} \textbf{\bibinfo{volume}{49}},
  \bibinfo{pages}{16223} (\bibinfo{year}{1994}),
  \urlprefix\url{https://link.aps.org/doi/10.1103/PhysRevB.49.16223}.

\bibitem[{\citenamefont{Kune\v{s} et~al.}(2010)\citenamefont{Kune\v{s}, Arita,
  Wissgott, Toschi, Ikeda, and Held}}]{wien2wannier}
\bibinfo{author}{\bibfnamefont{J.}~\bibnamefont{Kune\v{s}}},
  \bibinfo{author}{\bibfnamefont{R.}~\bibnamefont{Arita}},
  \bibinfo{author}{\bibfnamefont{P.}~\bibnamefont{Wissgott}},
  \bibinfo{author}{\bibfnamefont{A.}~\bibnamefont{Toschi}},
  \bibinfo{author}{\bibfnamefont{H.}~\bibnamefont{Ikeda}}, \bibnamefont{and}
  \bibinfo{author}{\bibfnamefont{K.}~\bibnamefont{Held}},
  \bibinfo{journal}{Computer Physics Communications}
  \textbf{\bibinfo{volume}{181}}, \bibinfo{pages}{1888} (\bibinfo{year}{2010}),
  ISSN \bibinfo{issn}{0010-4655},
  \urlprefix\url{https://www.sciencedirect.com/science/article/pii/S0010465510002948}.

\bibitem[{\citenamefont{Pizzi et~al.}(2020)\citenamefont{Pizzi, Vitale, Arita,
  Blügel, Freimuth, G{\'{e}}ranton, Gibertini, Gresch, Johnson, Koretsune
  et~al.}}]{wannier90}
\bibinfo{author}{\bibfnamefont{G.}~\bibnamefont{Pizzi}},
  \bibinfo{author}{\bibfnamefont{V.}~\bibnamefont{Vitale}},
  \bibinfo{author}{\bibfnamefont{R.}~\bibnamefont{Arita}},
  \bibinfo{author}{\bibfnamefont{S.}~\bibnamefont{Blügel}},
  \bibinfo{author}{\bibfnamefont{F.}~\bibnamefont{Freimuth}},
  \bibinfo{author}{\bibfnamefont{G.}~\bibnamefont{G{\'{e}}ranton}},
  \bibinfo{author}{\bibfnamefont{M.}~\bibnamefont{Gibertini}},
  \bibinfo{author}{\bibfnamefont{D.}~\bibnamefont{Gresch}},
  \bibinfo{author}{\bibfnamefont{C.}~\bibnamefont{Johnson}},
  \bibinfo{author}{\bibfnamefont{T.}~\bibnamefont{Koretsune}},
  \bibnamefont{et~al.}, \bibinfo{journal}{Journal of Physics: Condensed Matter}
  \textbf{\bibinfo{volume}{32}}, \bibinfo{pages}{165902}
  (\bibinfo{year}{2020}),
  \urlprefix\url{https://doi.org/10.1088/1361-648x/ab51ff}.

\bibitem[{\citenamefont{Mravlje et~al.}(2011)\citenamefont{Mravlje, Aichhorn,
  Miyake, Haule, Kotliar, and Georges}}]{Mravlje_Ru_dmft}
\bibinfo{author}{\bibfnamefont{J.}~\bibnamefont{Mravlje}},
  \bibinfo{author}{\bibfnamefont{M.}~\bibnamefont{Aichhorn}},
  \bibinfo{author}{\bibfnamefont{T.}~\bibnamefont{Miyake}},
  \bibinfo{author}{\bibfnamefont{K.}~\bibnamefont{Haule}},
  \bibinfo{author}{\bibfnamefont{G.}~\bibnamefont{Kotliar}}, \bibnamefont{and}
  \bibinfo{author}{\bibfnamefont{A.}~\bibnamefont{Georges}},
  \bibinfo{journal}{Phys. Rev. Lett.} \textbf{\bibinfo{volume}{106}},
  \bibinfo{pages}{096401} (\bibinfo{year}{2011}),
  \urlprefix\url{https://link.aps.org/doi/10.1103/PhysRevLett.106.096401}.

\bibitem[{\citenamefont{Kim et~al.}(2018)\citenamefont{Kim, Mravlje, Ferrero,
  Parcollet, and Georges}}]{Minjae_2018_PRL}
\bibinfo{author}{\bibfnamefont{M.}~\bibnamefont{Kim}},
  \bibinfo{author}{\bibfnamefont{J.}~\bibnamefont{Mravlje}},
  \bibinfo{author}{\bibfnamefont{M.}~\bibnamefont{Ferrero}},
  \bibinfo{author}{\bibfnamefont{O.}~\bibnamefont{Parcollet}},
  \bibnamefont{and} \bibinfo{author}{\bibfnamefont{A.}~\bibnamefont{Georges}},
  \bibinfo{journal}{Phys. Rev. Lett.} \textbf{\bibinfo{volume}{120}},
  \bibinfo{pages}{126401} (\bibinfo{year}{2018}),
  \urlprefix\url{https://link.aps.org/doi/10.1103/PhysRevLett.120.126401}.

\bibitem[{\citenamefont{Lee et~al.}(2020)\citenamefont{Lee, Kim, and
  Go}}]{Go_SRO_2020}
\bibinfo{author}{\bibfnamefont{H.~J.} \bibnamefont{Lee}},
  \bibinfo{author}{\bibfnamefont{C.~H.} \bibnamefont{Kim}}, \bibnamefont{and}
  \bibinfo{author}{\bibfnamefont{A.}~\bibnamefont{Go}}, \bibinfo{journal}{Phys.
  Rev. B} \textbf{\bibinfo{volume}{102}}, \bibinfo{pages}{195115}
  (\bibinfo{year}{2020}),
  \urlprefix\url{https://link.aps.org/doi/10.1103/PhysRevB.102.195115}.

\bibitem[{\citenamefont{Kugler et~al.}(2020)\citenamefont{Kugler, Zingl,
  Strand, Lee, von Delft, and Georges}}]{Kugler2020}
\bibinfo{author}{\bibfnamefont{F.~B.} \bibnamefont{Kugler}},
  \bibinfo{author}{\bibfnamefont{M.}~\bibnamefont{Zingl}},
  \bibinfo{author}{\bibfnamefont{H.~U.~R.} \bibnamefont{Strand}},
  \bibinfo{author}{\bibfnamefont{S.-S.~B.} \bibnamefont{Lee}},
  \bibinfo{author}{\bibfnamefont{J.}~\bibnamefont{von Delft}},
  \bibnamefont{and} \bibinfo{author}{\bibfnamefont{A.}~\bibnamefont{Georges}},
  \bibinfo{journal}{Phys. Rev. Lett.} \textbf{\bibinfo{volume}{124}},
  \bibinfo{pages}{016401} (\bibinfo{year}{2020}),
  \urlprefix\url{https://link.aps.org/doi/10.1103/PhysRevLett.124.016401}.

\bibitem[{\citenamefont{Cao et~al.}(2021)\citenamefont{Cao, Lu, Hansmann, and
  Haverkort}}]{Cao_2021}
\bibinfo{author}{\bibfnamefont{X.}~\bibnamefont{Cao}},
  \bibinfo{author}{\bibfnamefont{Y.}~\bibnamefont{Lu}},
  \bibinfo{author}{\bibfnamefont{P.}~\bibnamefont{Hansmann}}, \bibnamefont{and}
  \bibinfo{author}{\bibfnamefont{M.~W.} \bibnamefont{Haverkort}},
  \bibinfo{journal}{Phys. Rev. B} \textbf{\bibinfo{volume}{104}},
  \bibinfo{pages}{115119} (\bibinfo{year}{2021}),
  \urlprefix\url{https://link.aps.org/doi/10.1103/PhysRevB.104.115119}.

\bibitem[{\citenamefont{Kotliar et~al.}(2006)\citenamefont{Kotliar, Savrasov,
  Haule, Oudovenko, Parcollet, and Marianetti}}]{DMFT_RMP_2006}
\bibinfo{author}{\bibfnamefont{G.}~\bibnamefont{Kotliar}},
  \bibinfo{author}{\bibfnamefont{S.~Y.} \bibnamefont{Savrasov}},
  \bibinfo{author}{\bibfnamefont{K.}~\bibnamefont{Haule}},
  \bibinfo{author}{\bibfnamefont{V.~S.} \bibnamefont{Oudovenko}},
  \bibinfo{author}{\bibfnamefont{O.}~\bibnamefont{Parcollet}},
  \bibnamefont{and} \bibinfo{author}{\bibfnamefont{C.~A.}
  \bibnamefont{Marianetti}}, \bibinfo{journal}{Rev. Mod. Phys.}
  \textbf{\bibinfo{volume}{78}}, \bibinfo{pages}{865} (\bibinfo{year}{2006}),
  \urlprefix\url{https://link.aps.org/doi/10.1103/RevModPhys.78.865}.

\bibitem[{\citenamefont{Savrasov and Kotliar}(2004)}]{Savrasov2004}
\bibinfo{author}{\bibfnamefont{S.~Y.} \bibnamefont{Savrasov}} \bibnamefont{and}
  \bibinfo{author}{\bibfnamefont{G.}~\bibnamefont{Kotliar}},
  \bibinfo{journal}{Phys. Rev. B} \textbf{\bibinfo{volume}{69}},
  \bibinfo{pages}{245101} (\bibinfo{year}{2004}),
  \urlprefix\url{https://link.aps.org/doi/10.1103/PhysRevB.69.245101}.

\bibitem[{\citenamefont{Gilbert}(1975)}]{Gilbert1975}
\bibinfo{author}{\bibfnamefont{T.~L.} \bibnamefont{Gilbert}},
  \bibinfo{journal}{Phys. Rev. B} \textbf{\bibinfo{volume}{12}},
  \bibinfo{pages}{2111} (\bibinfo{year}{1975}),
  \urlprefix\url{https://link.aps.org/doi/10.1103/PhysRevB.12.2111}.

\bibitem[{\citenamefont{Chitra and Kotliar}(2000)}]{Chitra2000}
\bibinfo{author}{\bibfnamefont{R.}~\bibnamefont{Chitra}} \bibnamefont{and}
  \bibinfo{author}{\bibfnamefont{G.}~\bibnamefont{Kotliar}},
  \bibinfo{journal}{Phys. Rev. B} \textbf{\bibinfo{volume}{62}},
  \bibinfo{pages}{12715} (\bibinfo{year}{2000}),
  \urlprefix\url{https://link.aps.org/doi/10.1103/PhysRevB.62.12715}.

\bibitem[{\citenamefont{Adler et~al.}(2024)\citenamefont{Adler, Melnick, and
  Kotliar}}]{Portobello}
\bibinfo{author}{\bibfnamefont{R.}~\bibnamefont{Adler}},
  \bibinfo{author}{\bibfnamefont{C.}~\bibnamefont{Melnick}}, \bibnamefont{and}
  \bibinfo{author}{\bibfnamefont{G.}~\bibnamefont{Kotliar}},
  \bibinfo{journal}{Computer Physics Communications}
  \textbf{\bibinfo{volume}{294}}, \bibinfo{pages}{108907}
  (\bibinfo{year}{2024}), ISSN \bibinfo{issn}{0010-4655},
  \urlprefix\url{https://www.sciencedirect.com/science/article/pii/S0010465523002527}.

\bibitem[{\citenamefont{Lee et~al.}(2021)\citenamefont{Lee, Lanat\`a, Kim, and
  Kotliar}}]{Lee_PRX_2021}
\bibinfo{author}{\bibfnamefont{T.-H.} \bibnamefont{Lee}},
  \bibinfo{author}{\bibfnamefont{N.}~\bibnamefont{Lanat\`a}},
  \bibinfo{author}{\bibfnamefont{M.}~\bibnamefont{Kim}}, \bibnamefont{and}
  \bibinfo{author}{\bibfnamefont{G.}~\bibnamefont{Kotliar}},
  \bibinfo{journal}{Phys. Rev. X} \textbf{\bibinfo{volume}{11}},
  \bibinfo{pages}{041040} (\bibinfo{year}{2021}),
  \urlprefix\url{https://link.aps.org/doi/10.1103/PhysRevX.11.041040}.

\bibitem[{\citenamefont{Bauernfeind et~al.}(2017)\citenamefont{Bauernfeind,
  Zingl, Triebl, Aichhorn, and Evertz}}]{Bauernfeind2017}
\bibinfo{author}{\bibfnamefont{D.}~\bibnamefont{Bauernfeind}},
  \bibinfo{author}{\bibfnamefont{M.}~\bibnamefont{Zingl}},
  \bibinfo{author}{\bibfnamefont{R.}~\bibnamefont{Triebl}},
  \bibinfo{author}{\bibfnamefont{M.}~\bibnamefont{Aichhorn}}, \bibnamefont{and}
  \bibinfo{author}{\bibfnamefont{H.~G.} \bibnamefont{Evertz}},
  \bibinfo{journal}{Phys. Rev. X} \textbf{\bibinfo{volume}{7}},
  \bibinfo{pages}{031013} (\bibinfo{year}{2017}),
  \urlprefix\url{https://link.aps.org/doi/10.1103/PhysRevX.7.031013}.

\end{thebibliography}

\end{document}